%% file: main.tex
\pdfoutput=1
\documentclass[12pt,a4paper]{article}
\usepackage{ifthen} 
\newboolean{pdflatex}
\setboolean{pdflatex}{true} 

\newboolean{articletitles}
\setboolean{articletitles}{true} 

\newboolean{uprightparticles}
\setboolean{uprightparticles}{false} 

\def\paperauthors{
F.~Alessio$^{1}$\lhcborcid{0000-0001-5317-1098},
S.~Barsuk$^{2}$\lhcborcid{0000-0002-0898-6551},
C.~Beigbeder$^{2}$\lhcborcid{0009-0003-6086-7087},
A.~Bellavista$^{3,4}$\lhcborcid{0009-0009-3723-834X},
O.~Bezshyyko$^{2}$\lhcborcid{0000-0001-7106-5213},
I.~Boiaryntseva$^{2,5}$\lhcborcid{0000-0002-5219-6127},
C.~Bourgeois$^{2}$\lhcborcid{0009-0007-4215-4955},
A.~Boyarintsev$^{5}$\lhcborcid{0000-0001-9252-0430},
M.~Briere$^{2}$\lhcborcid{0009-0005-9050-9728},
L.~Burmistrov$^{6}$,
A.~Carbone$^{3,4}$\lhcborcid{0000-0002-7045-2243},
G.~Cavallero$^{7}$\lhcborcid{0000-0002-8342-7047},
V.~Chaumat$^{2}$\lhcborcid{0009-0003-4363-3318},
S.~Cholak$^{8,2}$\lhcborcid{0000-0001-8091-4766},
D.~Douillet$^{2}$,
O.~Duarte$^{2}$\lhcborcid{0009-0000-2233-3162},
P.~Durante$^{1}$\lhcborcid{0000-0002-1204-2270},
F.~Ferrari$^{3,4}$\lhcborcid{0000-0002-3721-4585},
E.~Franzoso$^{7}$\lhcborcid{0000-0003-2130-1593},
C.~Gaspar$^{1}$\lhcborcid{0000-0002-8009-1509},
L.~Golinka-Bezshyyko$^{2}$\lhcborcid{0000-0002-0613-5374},
L.A.~Granado~Cardoso$^{1}$\lhcborcid{0000-0003-2868-2173},
E.~Graverini$^{8}$\lhcborcid{0000-0003-4647-6429},
D.~Hohov$^{2}$\lhcborcid{0000-0002-4760-1597},
G.~Iaquaniello$^{2}$\lhcborcid{0009-0007-8161-8652},
V.~Kushnir$^{9}$\lhcborcid{0000-0003-2907-1323},
R.~Matev$^{1}$\lhcborcid{0000-0001-8713-6119},
P.~Mayencourt$^{8}$\lhcborcid{0000-0002-8210-1256},
V.~Mytrochenko$^{9}$\lhcborcid{ 0000-0002-3002-7402},
V.~Orlov$^{2}$\lhcborcid{0009-0001-0028-9647},
V.~Puill$^{2}$\lhcborcid{0000-0003-0806-7149},
D.~Reynet$^{2}$,
P.~Robbe$^{2}$\lhcborcid{0000-0002-0656-9033},
P.~Rosier$^{2}$\lhcborcid{0000-0002-2105-6308},
L.~Shchutska$^{8}$\lhcborcid{0000-0003-0700-5448},
Y.~Song$^{8}$\lhcborcid{0000-0003-0256-4320},
E.~Spedicato$^{3,4}$\lhcborcid{0000-0002-4950-6665},
L.~Toscano$^{10}$\lhcborcid{0009-0007-5613-6520},
M.~van Dijk$^{1}$\lhcborcid{0000-0003-2538-5798},
A.~Villa$^{8}$\lhcborcid{0000-0002-9392-6157},
G.~Vouters$^{11}$\lhcborcid{0009-0008-3292-2209},
V.~Yeroshenko$^{2}$\lhcborcid{0000-0002-8771-0579},
V.~Zhovkovska$^{12}$\lhcborcid{0000-0002-9812-4508}
\\
{\bigskip \footnotesize \it
$^{1}$European Organization for Nuclear Research (CERN), Geneva, Switzerland\\
$^{2}$Laboratoire de Physique des 2 Infinis Irène Joliot-Curie, IJCLab, Orsay, France\\
$^{3}$Sezione INFN di Bologna, Bologna, Italy\\
$^{4}$Università di Bologna, Bologna, Italy\\
$^{5}$Institute for Scintillation Materials of the NAS of Ukraine, Kharkiv, Ukraine\\
$^{6}$Département de Physique Nucléaire et Corpusculaire, Université de Genève, Genève, Switzerland\\
$^{7}$Sezione INFN di Ferrara, Ferrara, Italy\\
$^{8}$Institute of Physics, École Polytechnique Fédérale de Lausanne (EPFL), Lausanne, Switzerland\\
$^{9}$Kharkiv Institute of Physics and Technology, Kharkiv, Ukraine\\
$^{10}$Fakultät Physik, Technische Universität Dortmund, Dortmund, Germany\\
$^{11}$Univ. Savoie Mont Blanc, CNRS, IN2P3-LAPP, Annecy, France\\
$^{12}$Department of Physics, University of Warwick, Coventry, United Kingdom\\
}
}

\def\paperasciititle{Operation and performance of the Probe for Luminosity Measurement at LHCb} 
\def\papertitle{Operation and performance of the Probe for Luminosity Measurement at LHCb} 
\def\paperkeywords{{High Energy Physics}, {LHCb}} 
\def\papercopyright{\the\year\ CERN for the benefit of the LHCb collaboration} 
\def\paperlicence{CC BY 4.0 licence}
\def\paperlicenceurl{https://creativecommons.org/licenses/by/4.0/}

\newif\ifEnableSectionTOCLinks
\EnableSectionTOCLinksfalse 

\input{preamble}
\usepackage{longtable} 

\begin{document}

\renewcommand{\thefootnote}{\fnsymbol{footnote}}
\setcounter{footnote}{1}


\input{title-LHCb-PAPER}


\renewcommand{\thefootnote}{\arabic{footnote}}
\setcounter{footnote}{0}


\cleardoublepage


\pagestyle{plain} 
\setcounter{page}{1}
\pagenumbering{arabic}


\input{body.tex}

\addcontentsline{toc}{section}{References}
\bibliographystyle{LHCb}
\bibliography{main,standard,LHCb-PAPER,LHCb-CONF,LHCb-DP,LHCb-TDR}
 
\end{document}

%% file: preamble.tex
\usepackage[top=1in, bottom=1.25in, left=1in, right=1in]{geometry}

\usepackage{siunitx}
\usepackage{microtype}
\usepackage{lineno}  
\usepackage{xspace} 
\usepackage{booktabs}
\usepackage{caption} 

\usepackage{graphicx}  
\usepackage{color}
\usepackage{colortbl}
\graphicspath{{./figs/}} 

\usepackage{amsmath} 
\usepackage{amssymb}
\usepackage{amsfonts}
\usepackage{upgreek} 

\usepackage{booktabs}
\newcommand*\patchAmsMathEnvironmentForLineno[1]{%
\expandafter\let\csname old#1\expandafter\endcsname\csname #1\endcsname
\expandafter\let\csname oldend#1\expandafter\endcsname\csname
end#1\endcsname
 \renewenvironment{#1}%
   {\linenomath\csname old#1\endcsname}%
   {\csname oldend#1\endcsname\endlinenomath}%
}
\newcommand*\patchBothAmsMathEnvironmentsForLineno[1]{%
  \patchAmsMathEnvironmentForLineno{#1}%
  \patchAmsMathEnvironmentForLineno{#1*}%
}
\AtBeginDocument{%
\patchBothAmsMathEnvironmentsForLineno{equation}%
\patchBothAmsMathEnvironmentsForLineno{align}%
\patchBothAmsMathEnvironmentsForLineno{flalign}%
\patchBothAmsMathEnvironmentsForLineno{alignat}%
\patchBothAmsMathEnvironmentsForLineno{gather}%
\patchBothAmsMathEnvironmentsForLineno{multline}%
\patchBothAmsMathEnvironmentsForLineno{eqnarray}%
}

\usepackage[pdftex,
            pdftitle={\paperasciititle},
            pdfkeywords={\paperkeywords},
            ]{hyperref}
\usepackage{hyperxmp}
\usepackage{cleveref}

\usepackage[colorinlistoftodos,textsize=scriptsize]{todonotes}

\usepackage[bottom,flushmargin,hang,multiple]{footmisc}

\usepackage[all]{hypcap} 

\input{lhcb-symbols-def} 

\hypersetup{
  colorlinks   = true, 
  urlcolor     = blue, 
  linkcolor    = blue, 
  citecolor    = red   
}

\ifEnableSectionTOCLinks
    \usepackage[explicit]{titlesec} 
    
    \let\oldcontentsline\contentsline
    \renewcommand

    \titleformat{\section}{\normalfont\Large\bf}{\hyperlink{tocsection.\thesection}{{\thesection} \parbox[t]{\dimexpr\textwidth-1pc}{#1}}}{1pc}{}

    \titleformat{\subsection}{\normalfont\bf}{\hyperlink{tocsubsection.\thesubsection}{{\thesubsection} \parbox[t]{\dimexpr\textwidth-1pc}{#1}}}{1pc}{}

    \titleformat{name=\section,numberless}[display]{}{}{0pt}{\normalfont\Huge\bfseries #1}
\fi

\usepackage{cite} 
\usepackage{mciteplus}

%% file: lhcb-symbols-def.tex
\usepackage{xspace} 
\usepackage{upgreek}

\def\MagUp {\mbox{\em Mag\kern -0.05em Up}\xspace}

\def\hlttwo {HLT2\xspace}

\ifthenelse{\boolean{uprightparticles}}%
{

 \def\PDelta      {\ensuremath{\Delta}\xspace}                 
 \def\PXi         {\ensuremath{\Xi}\xspace}                 
 \def\PLambda     {\ensuremath{\Lambda}\xspace}                 
 \def\PSigma      {\ensuremath{\Sigma}\xspace}                 
 \def\POmega      {\ensuremath{\Omega}\xspace}                 
 \def\PUpsilon    {\ensuremath{\Upsilon}\xspace}
 \let\oldPi\Pi
 \def\PPi         {\ensuremath{\oldPi}\xspace}

 \def\PB      {\ensuremath{\mathrm{B}}\xspace}                 
 \def\PD      {\ensuremath{\mathrm{D}}\xspace}                 

 \def\PK      {\ensuremath{\mathrm{K}}\xspace}                 
 \def\Pp      {\ensuremath{\mathrm{p}}\xspace}                 

 \def\Ps      {\ensuremath{\mathrm{s}}\xspace}

 \def\thebaroffset{0.0em}
}
{

 \mathchardef\PDelta="7101
 \mathchardef\PXi="7104
 \mathchardef\PLambda="7103
 \mathchardef\PSigma="7106
 \mathchardef\POmega="710A
 \mathchardef\PUpsilon="7107
 \mathchardef\PPi="7105
 \def\PB      {\ensuremath{B}\xspace}                 
 \def\PD      {\ensuremath{D}\xspace}                 

 \def\PK      {\ensuremath{K}\xspace}                 
 \def\Pp      {\ensuremath{p}\xspace}                 

 \def\Ps      {\ensuremath{s}\xspace}

 \def\thebaroffset{0.18em}
}
\newcommand{\offsetoverline}[2][\thebaroffset]{\kern #1\overline{\kern -#1 #2}}%

\makeatletter
\ifcase \@ptsize \relax
  \newcommand{\miniscule}{\@setfontsize\miniscule{4}{5}}
\or
  \newcommand{\miniscule}{\@setfontsize\miniscule{5}{6}}
\or
  \newcommand{\miniscule}{\@setfontsize\miniscule{5}{6}}
\fi
\makeatother

\DeclareRobustCommand{\optbar}[1]{\shortstack{{\miniscule (\rule[.5ex]{1.25em}{.18mm})}
  \\ [-.7ex] $#1$}}

\def\squark    {{\ensuremath{\Ps}}\xspace}

\def\KorKbar {\kern \thebaroffset\optbar{\kern -\thebaroffset \PK}{}\xspace}

\def\D       {{\ensuremath{\PD}}\xspace}

\def\DorDbar {\kern \thebaroffset\optbar{\kern -\thebaroffset \PD}\xspace}

\def\Dp      {{\ensuremath{\D^+}}\xspace}
\def\Dm      {{\ensuremath{\D^-}}\xspace}

\def\DpDm    {\ensuremath{\Dp {\kern -0.16em \Dm}}\xspace}

\def\B       {{\ensuremath{\PB}}\xspace}

\def\BorBbar {\kern \thebaroffset\optbar{\kern -\thebaroffset \PB}\xspace}

\def\Bd      {{\ensuremath{\B^0}}\xspace}

\def\BdorBdbar {\kern \thebaroffset\optbar{\kern -\thebaroffset \Bd}\xspace}

\def\Bs      {{\ensuremath{\B^0_\squark}}\xspace}

\def\BsorBsbar {\kern \thebaroffset\optbar{\kern -\thebaroffset \Bs}\xspace}

\def\Y#1S{\ensuremath{\PUpsilon{(#1S)}}\xspace}

\def\proton      {{\ensuremath{\Pp}}\xspace}

\def\LorLbar     {\kern \thebaroffset\optbar{\kern -\thebaroffset \PLambda}\xspace}

\def\AT#1     {\ensuremath{A_{\mathrm{T}}^{#1}}\xspace}           

\def\C#1      {\ensuremath{\mathcal{C}_{#1}}\xspace}                       
\def\Cp#1     {\ensuremath{\mathcal{C}_{#1}^{'}}\xspace}                    
\def\Ceff#1   {\ensuremath{\mathcal{C}_{#1}^{\mathrm{(eff)}}}\xspace}        
\def\Cpeff#1  {\ensuremath{\mathcal{C}_{#1}^{'\mathrm{(eff)}}}\xspace}       
\def\Ope#1    {\ensuremath{\mathcal{O}_{#1}}\xspace}                       
\def\Opep#1   {\ensuremath{\mathcal{O}_{#1}^{'}}\xspace}                    

\newcommand{\aunit}[1]{\ensuremath{\text{\,#1}}}       

\newcommand{\tev}{\aunit{Te\kern -0.1em V}\xspace}
\newcommand{\gev}{\aunit{Ge\kern -0.1em V}\xspace}
\newcommand{\mev}{\aunit{Me\kern -0.1em V}\xspace}
\newcommand{\kev}{\aunit{ke\kern -0.1em V}\xspace}
\newcommand{\ev}{\aunit{e\kern -0.1em V}\xspace}
 
\newcommand{\mevc}{\ensuremath{\aunit{Me\kern -0.1em V\!/}c}\xspace}
\newcommand{\gevc}{\ensuremath{\aunit{Ge\kern -0.1em V\!/}c}\xspace}
\newcommand{\mevcc}{\ensuremath{\aunit{Me\kern -0.1em V\!/}c^2}\xspace}
\newcommand{\gevcc}{\ensuremath{\aunit{Ge\kern -0.1em V\!/}c^2}\xspace}

\def\m    {\aunit{m}\xspace}

\def\fb   {\ensuremath{\aunit{fb}}\xspace}
\def\invfb   {\ensuremath{\fb^{-1}}\xspace}

\def\gsim{{~\raise.15em\hbox{$>$}\kern-.85em
          \lower.35em\hbox{$\sim$}~}\xspace}
\def\lsim{{~\raise.15em\hbox{$<$}\kern-.85em
          \lower.35em\hbox{$\sim$}~}\xspace}

\def\tell1  {TELL1\xspace}
\def\ukl1   {UKL1\xspace}

\newcommand{\eg}{\mbox{\itshape e.g.}\xspace}
\newcommand{\ie}{\mbox{\itshape i.e.}\xspace}

\newcommand{\lhcborcid}[1]{\href{https://orcid.org/#1}{\hspace*{0.1em}\raisebox{-0.45ex}{\includegraphics[width=1em]{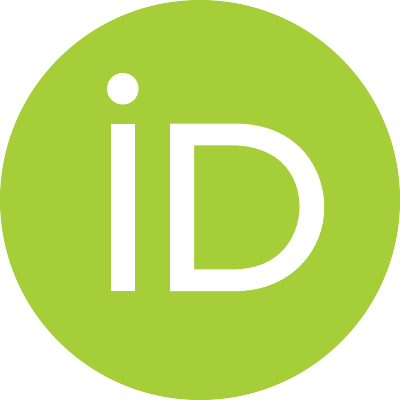}}}}

\def\sigmavis {\ensuremath{\sigma_{\mathrm{vis}}}\xspace}
\def\mupeak  {\ensuremath{\mu_{\mathrm{bkg-sub}}^{\mathrm{peak}}}\xspace}

\def\mubkgsub {\ensuremath{\mu_{\mathrm{bkg-sub}}}\xspace}
\def\mubb {\ensuremath{\mu_{\mathrm{bb}}}\xspace}
\def\mube {\ensuremath{\mu_{\mathrm{be}}}\xspace}
\def\mueb {\ensuremath{\mu_{\mathrm{eb}}}\xspace}
\def\muee {\ensuremath{\mu_{\mathrm{ee}}}\xspace}
\def\avgNbb {\ensuremath{\langle N_{\mathrm{bb}} \rangle}\xspace}
\def\avgNbe {\ensuremath{\langle N_{\mathrm{be}} \rangle}\xspace}
\def\avgNeb {\ensuremath{\langle N_{\mathrm{eb}} \rangle}\xspace}
\def\bb {\ensuremath{\mathrm{bb}}\xspace}
\def\be {\ensuremath{\mathrm{be}}\xspace}
\def\eb {\ensuremath{\mathrm{eb}}\xspace}
\def\ee {\ensuremath{\mathrm{ee}}\xspace}
\def\nbb {\ensuremath{n_{\mathrm{bb}}}\xspace}
\def\LumiTwentyFour {\ensuremath{9.56}\xspace}
\def\LumiErrTwentyFour {\ensuremath{0.39}\xspace}
\def\LumiErrRelTwentyFour {\ensuremath{4.0}\xspace}
\def\LumiTwentyFive {\ensuremath{11.81}\xspace}
\def\LumiErrTwentyFive {\ensuremath{0.48}\xspace}

\def\LumiTwentySix {\ensuremath{5.34}\xspace}
\def\LumiErrTwentySix {\ensuremath{0.22}\xspace}

\def\LumiRunThree {\ensuremath{26.71}\xspace}
\def\LumiErrRunThree {\ensuremath{1.07}\xspace}


%% file: title-LHCb-PAPER.tex

\begin{titlepage}
\pagenumbering{roman}

\vspace*{-1.5cm}
\centerline{\large EUROPEAN ORGANIZATION FOR NUCLEAR RESEARCH (CERN)}
\vspace*{1.5cm}
\noindent
\begin{tabular*}{\linewidth}{lc@{\extracolsep{\fill}}r@{\extracolsep{0pt}}}
\ifthenelse{\boolean{pdflatex}}
{\vspace*{-1.5cm}\mbox{\!\!\!\includegraphics[width=.14\textwidth]{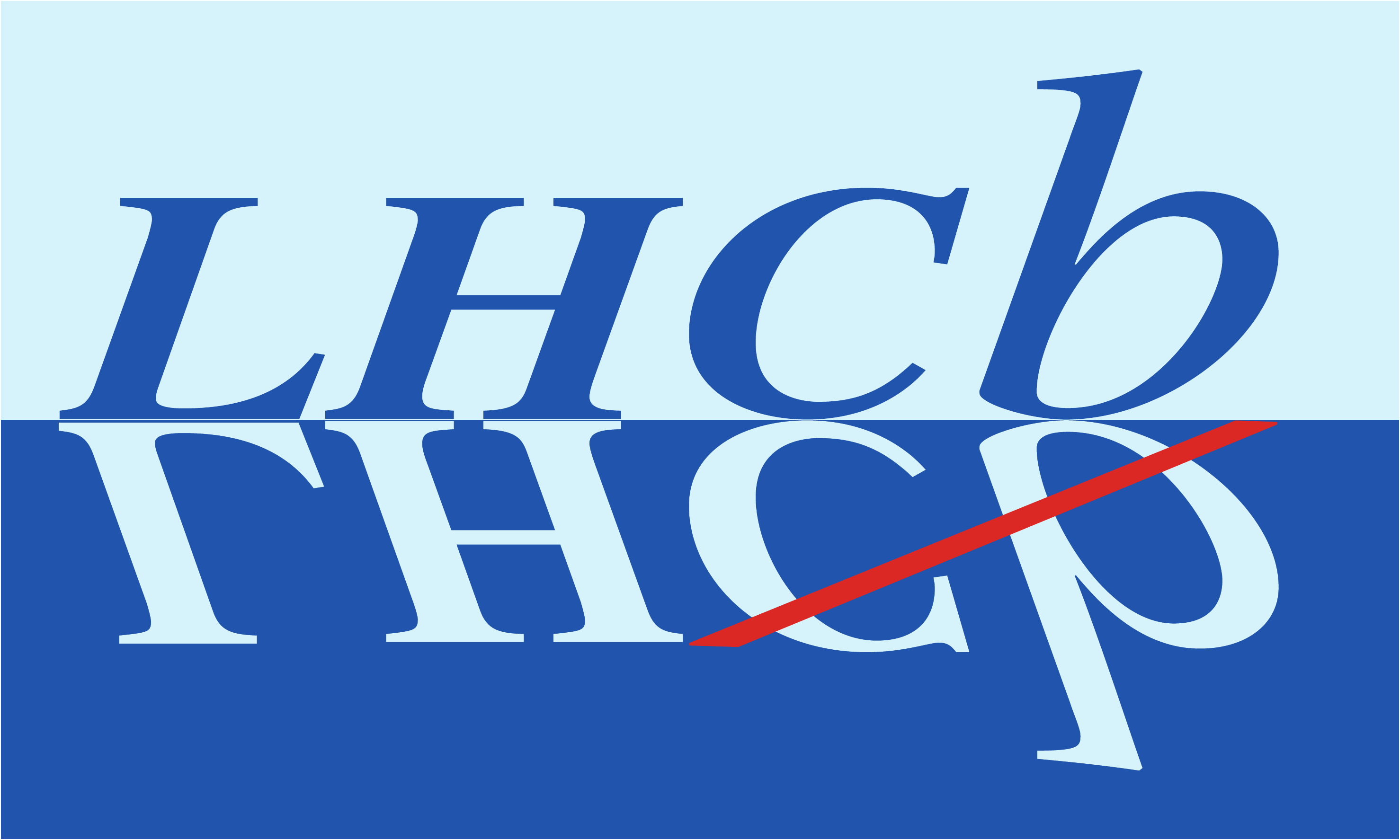}} & &}%
{\vspace*{-1.2cm}\mbox{\!\!\!\includegraphics[width=.12\textwidth]{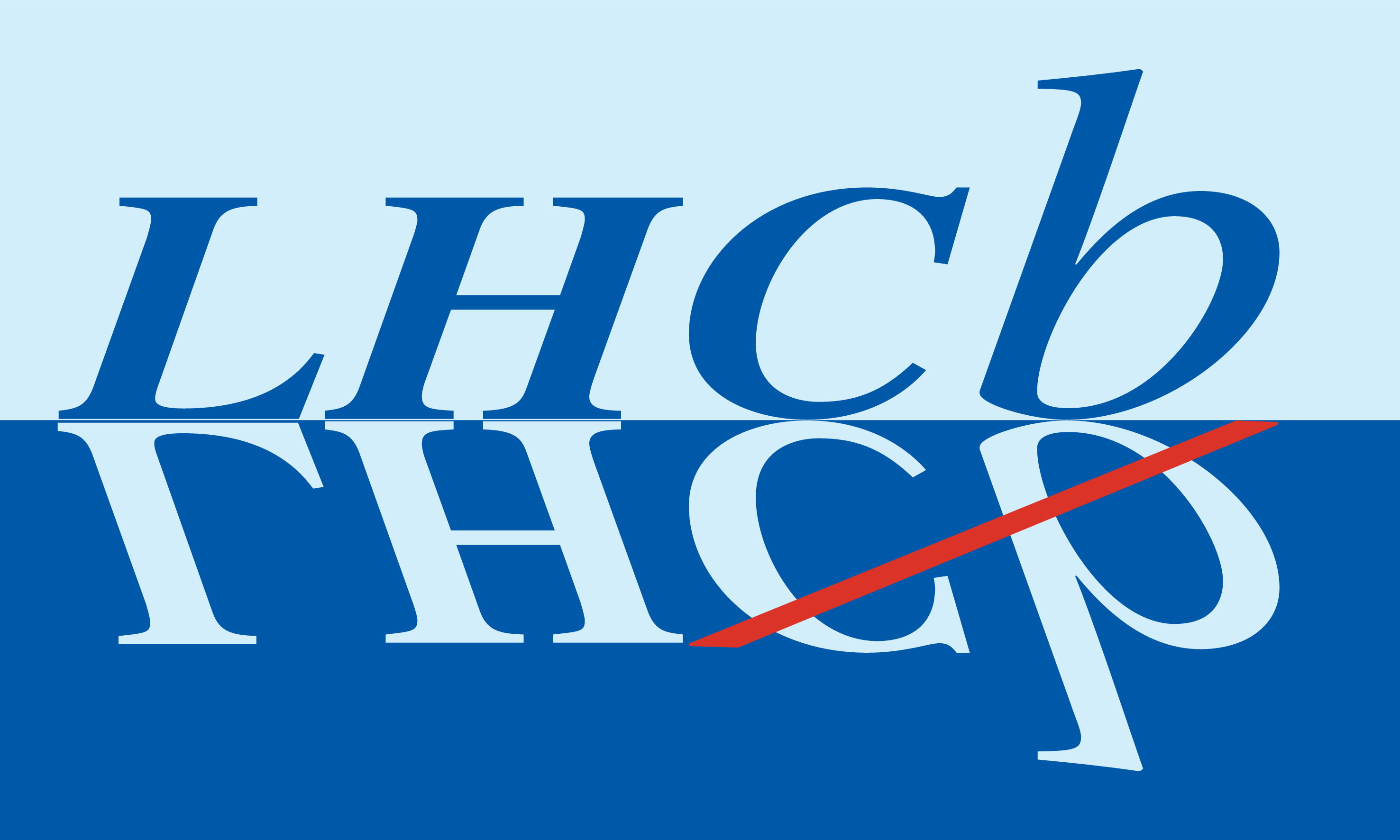}} & &}%
\\
 & & LHCb-DP-2026-001 \\  
 & & \today \\ 
 & & \\
\end{tabular*}

\vspace*{2.0cm}

{\normalfont\bfseries\boldmath\huge
\begin{center}
  \papertitle 
\end{center}
}

\vspace*{1.2cm}

\begin{center}
\paperauthors
\end{center}

\begin{abstract}
  \noindent
The Probe for LUminosity MEasurement (PLUME) detector is a dedicated luminosity monitor at the LHCb interaction point. It is a hodoscope comprising 48 Hamamatsu R760 photomultiplier tubes that detect Cherenkov light produced by particles travelling in the direction opposite to the LHCb spectrometer acceptance. PLUME provides real-time, bunch-by-bunch luminosity measurements during LHCb data taking and serves as the primary detector for controlling the luminosity levelling process at LHCb. In addition, its data are used offline, together with other dedicated counters from LHCb sub-systems, to determine the integrated luminosity delivered to the experiment. This paper reports on the detector’s operational performance during the 2024--2026 data-taking period of Run 3. The integrated luminosity recorded by the LHCb experiment in \(pp\) collisions at \(\sqrt{s}=13.6\)~\tev during this period, as measured by the PLUME online luminosity counters for detector performance monitoring, amounts to \(\mathcal{L} = (\LumiRunThree \pm \LumiErrRunThree)\)~\invfb. The luminosity values used in physics analyses are determined separately through dedicated offline calibrations based on van der Meer scans and Beam Gas Imaging techniques, and will be reported in a dedicated publication, as they are beyond the scope of this paper.
  
\end{abstract}

\vspace*{2.0cm}

\vspace{\fill}
\begin{center}
  Published in JINST 21 (2026) P09003
\end{center}

{\footnotesize 
\centerline{\copyright~\papercopyright. \href{\paperlicenceurl}{\paperlicence}.}}
\vspace*{2mm}

\end{titlepage}


\newpage
\setcounter{page}{2}
\mbox{~}
%
%
%
%

%% file: body.tex
\section{Introduction}
\label{sec:intro}

The Large Hadron Collider beauty (LHCb) experiment is one of the nine particle-physics experiments located along the Large Hadron Collider (LHC) at CERN~\cite{LHCb-DP-2008-001}.
LHCb is dedicated to flavour physics, focusing on precise studies of charge–parity (CP) violation and other rare processes involving beauty and charm hadrons produced in proton–proton collisions.

In Run 3, the upgraded LHCb detector has operated at luminosities approximately five times higher than those achieved in previous runs~\cite{LHCb-DP-2022-002}.
In particle colliders, luminosity is a key machine parameter that requires continuous monitoring.
The forward acceptance of the LHCb detector leads to high detector occupancies, requiring careful optimization of the instantaneous luminosity to maintain overall detector performance. 
As a result, LHCb has historically operated at lower luminosities than ATLAS and CMS, although future detector upgrades are designed to extend this capability significantly.
For this reason, the instantaneous luminosity delivered to LHCb is stabilised through a dedicated luminosity-levelling procedure~\cite{Muratori:2014ija}.
A precise, real-time determination of the instantaneous luminosity is therefore essential for the stable operation and data-taking performance of the experiment.

\begin{figure}[!htbp]
\centering
\includegraphics[width=0.6\linewidth]{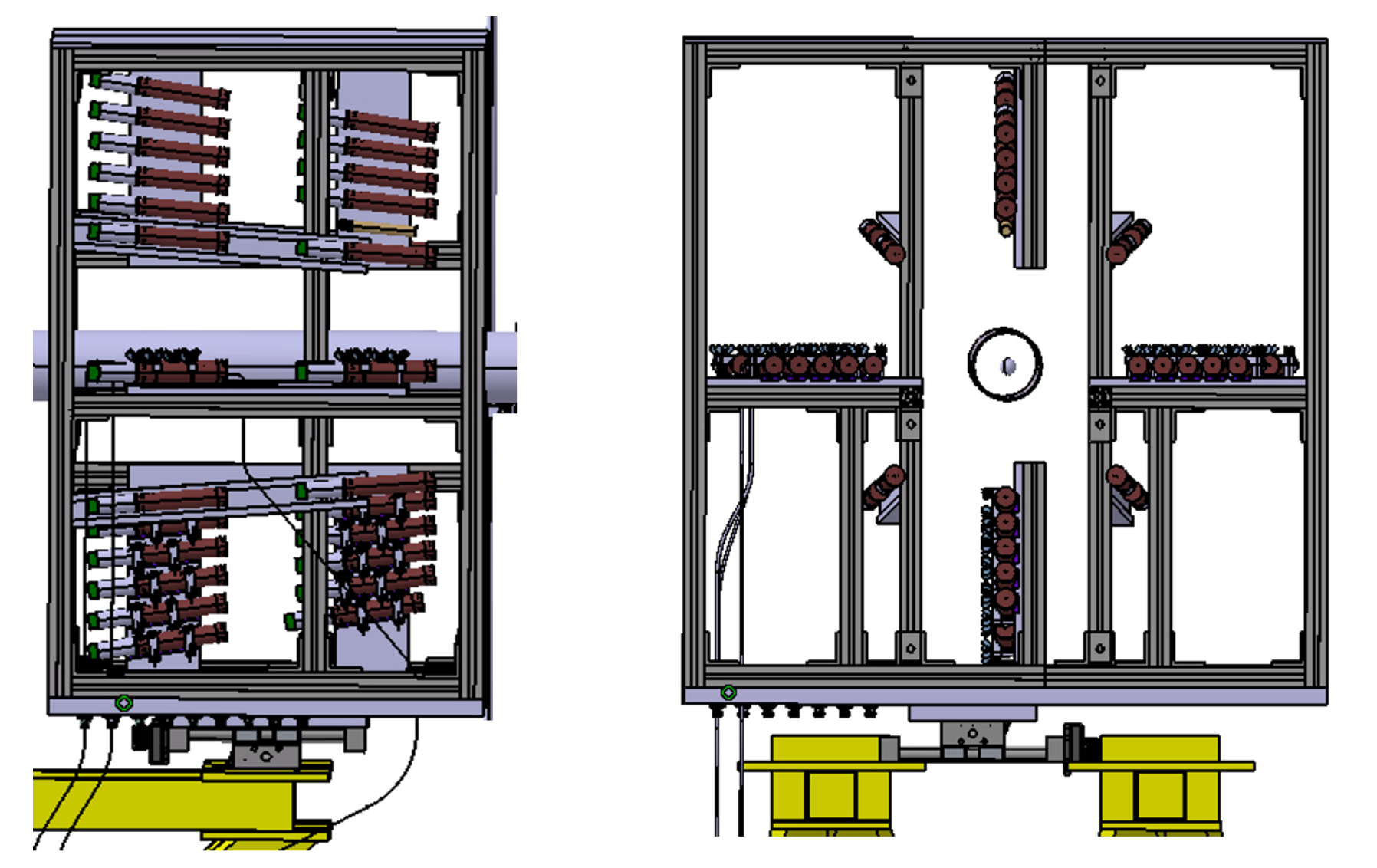}
\includegraphics[width=0.39\linewidth]{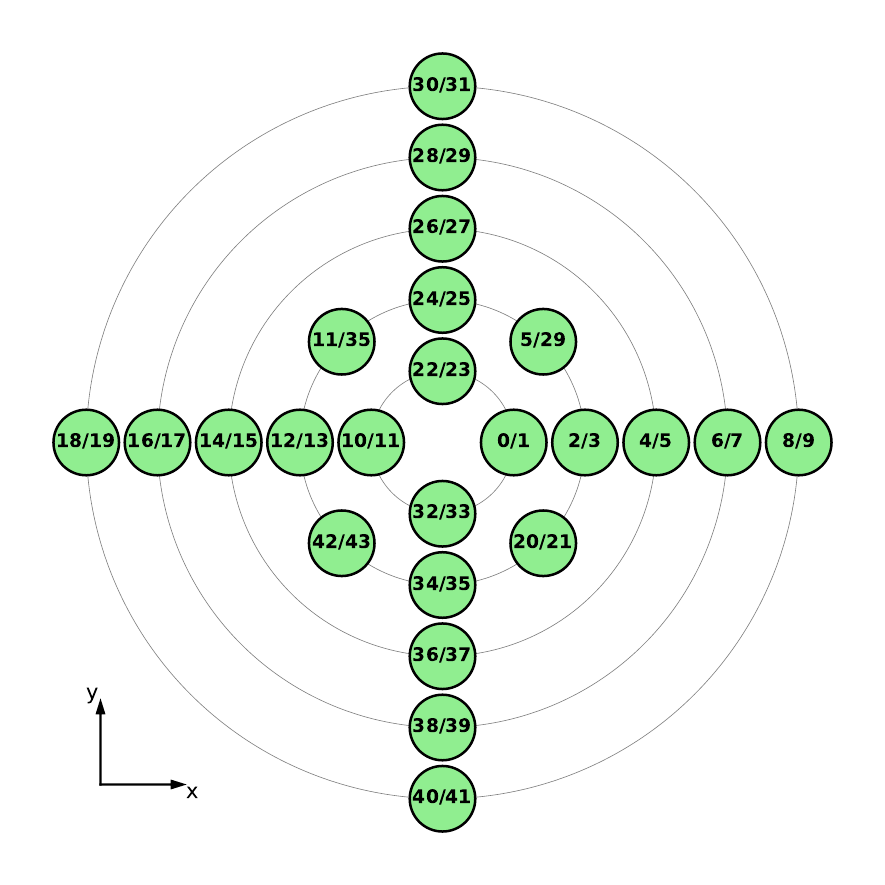}
\caption{\label{fig:PLUME} (Left) PLUME detector arrangement in the $y$–$z$ (left) and $x$–$y$ (right) planes. Cable and fibre routing for one elementary detection module is also shown. (Right) PLUME detector PMT numbering scheme.}
\end{figure}

\begin{figure}[!htbp]
\centering
\includegraphics[width=0.8\linewidth]{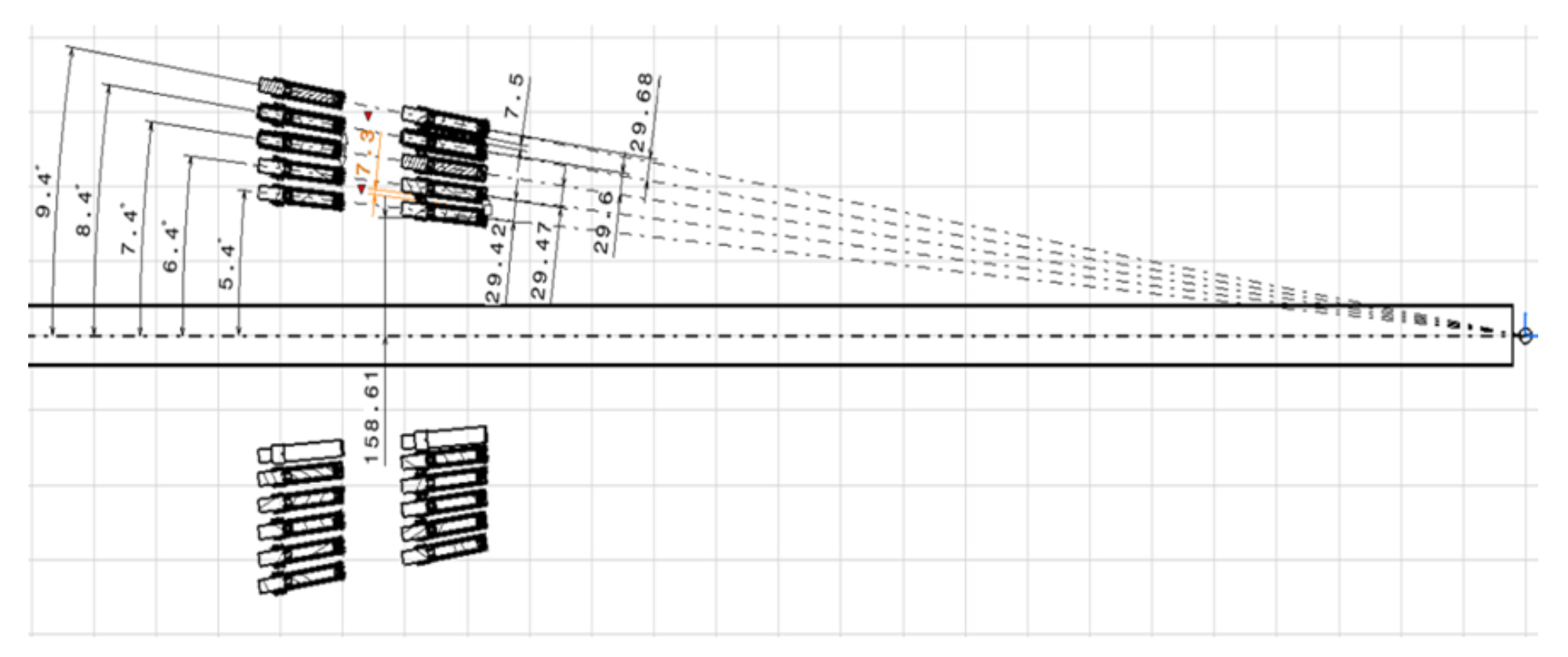}
\caption{\label{fig:PLUME_angular position} Angular positions of the PLUME elementary detection modules in the $y$–$z$ plane with respect to the nominal pp collision vertex.}
\end{figure}

The Probe for LUminosity MEasurement (PLUME), the luminometer of the upgraded LHCb detector, is designed to precisely monitor the LHC beam conditions at the LHCb interaction point, providing low-latency and continuous luminosity measurements to the leveling algorithm.
PLUME is a hodoscope composed of 48 head-on photomultiplier tubes (PMTs) of type R760 from Hamamatsu; it measures luminosity by detecting Cherenkov light produced in fused silica by charged particles emerging from the collision region.
A detailed characterisation of the PMTs and their performance has been done, as described in Ref.~\cite{Bellavista_2026}.
The overall detector layout is shown in the left panel of figure~\ref{fig:PLUME}, while the adopted PMT numbering scheme is displayed in the right panel.
The detection modules are arranged in a projective geometry, forming a two-layer hodoscope with a cross-shaped configuration around the beam pipe, as depicted in figure~\ref{fig:PLUME_angular position}.
The PLUME detector is installed between $z=-1485$ mm and $z=-2085$ mm along the beam axis, covering a pseudorapidity range of $-2.4 < \eta < -3.1$.

In addition to luminosity measurements, four PMTs are dedicated to the determination of the beam-phase timing.
Since the LHC clock is distributed to the major experiments via fibres that are not temperature-stabilised, variations in the beam phase are expected between day and night (at the level of hundreds of picoseconds) and across seasons (up to a few nanoseconds).
Such variations can lead to efficiency losses due to imperfect time alignment among the various sub-detectors.
It is therefore essential to monitor and correct for this effect.
To monitor these variations, two button electrode beam pick-ups (BPTX)~\cite{BPTX,Alessio:2011yya}, part of the LHC beam instrumentation and located in the vicinity of the LHCb interaction region, are used.
PLUME provides an independent measurement, with a precision of 8 ps, as shown by variations from consecutive measurements.

The paper is organised as follows. 
After this brief introduction, Section~\ref{sec:lumi_measurement} describes how the luminosity is measured and the LHC clock phase is monitored.
Section~\ref{sec:detector} describes the adopted detector technology, the front-end and back-end boards used for the PLUME readout, and the monitoring system employed for gain adjustment; it also outlines the development of the firmware and slow-control systems required for online luminosity and beam-phase timing measurements.
The detector operations during the LHC Run 3, as well as the corresponding performance, are discussed in Section~\ref{sec:performance}.
Finally, conclusions are given in Section~\ref{sec:conclusions}.

\section{Luminosity and LHC clock phase monitoring}
\label{sec:lumi_measurement}

This section describes the monitoring capabilities of the system with respect to luminosity and the LHC clock phase.
In particular, it outlines the methods adopted to measure the instantaneous luminosity during data taking and to monitor the relative phase of the LHC clock.
These two aspects are closely related to the operational performance of the detector and provide essential information for both data quality assessment and stable running conditions.

\subsection{Luminosity determination}
Luminosity is determined from the average number of interactions per bunch crossing observed by a given detector (\eg a single PLUME PMT), denoted as $\mu$.
The subscript highlights that the value of $\mu$ is detector-dependent, as it corresponds to the average number of detected interactions rather than the total number of proton–proton interactions.

The PLUME detector adopts the logZero method, which relies on the relation \mbox{$P(0) = e^{-\mu}$}, valid for a Poisson process.
This leads to the expression \mbox{$\mu = -\log P(0)$}.
The probability $P(0)$ is estimated as $N_{0}/N$, where $N_{0}$ denotes the number of detector-empty events, \ie events in which the measured signal is consistent with zero (typically a signal below the chosen threshold), out of a total of $N$ events.
Since $N_{0}$ follows a binomial distribution and the logarithm is a non-linear function, a second-order expansion in $(N_{0} - \langle N_{0} \rangle)$ leads to a bias correction term, yielding the estimator
\begin{equation}
    \label{eq:mu_def}
    \mu =
    -\log\left(\frac{N_{0}}{N}\right)
    - \frac{1}{2}\left(\frac{1}{N_{0}} - \frac{1}{N}\right).
\end{equation}
A detailed derivation of the second-order correction can be found in Ref.~\cite{LHCb-TDR-022}.

The LHCb detector is also equipped with a 20 cm-long storage cell, known as SMOG2~\cite{LHCb-DP-2024-002}, which allows the experiment to operate in fixed-target mode concurrently with nominal data taking. 
For a precise luminosity measurement, the contribution from interactions occurring in the SMOG2 cell must be subtracted from those taking place in the nominal interaction region.
The background subtraction is performed for each PLUME PMT according to
\begin{equation}
    \label{eq:mu_bkg_sub}
    \mubkgsub =
    \mubb
    - \frac{\avgNbb}{\avgNbe}\,\mube
    - \frac{\avgNbb}{\avgNeb}\,\mueb
    + \muee ,
\end{equation}
where $\mu_x$, with $x = \bb,\,\be,\,\eb,\,\mathrm{or}\,\ee$, denotes the average number of detected collisions in beam--beam, beam--empty, empty--beam, and empty--empty crossings, respectively. 
Here, beam--beam corresponds to crossings with bunches in both beams, beam--empty (empty--beam) to crossings with a bunch in beam 1 (beam 2) only, and empty--empty to crossings with no bunches in either beam.
The beam--beam term includes both $pp$ or $\mathrm{PbPb}$ interactions and collisions between beam particles and the injected gas.
The quantities $\mube$ and $\mueb$ serve as proxies for the beam--gas and gas--beam contributions, respectively.
Since the number of particles per bunch in beam--empty and empty--beam crossings decay differently from those in beam--beam crossings, correction factors are applied to $\mube$ and $\mueb$.
These factors, \avgNbb, \avgNbe, and \avgNeb, are derived from LHC measurements and used as external parameters in the luminosity determination.
Finally, the \muee term accounts for interactions due to the ghost charge, which are assumed to be uniformly distributed across all bunch crossings; their population are not corrected for as they are below the sensitivities of the dedicated LHC instruments.

The relationship between \mubkgsub and the instantaneous luminosity $\mathcal{L}_\mathrm{inst}$ is
\begin{equation} \label{eq:lumi}
    \mathcal{L}_\mathrm{inst} = \nbb\, f \, \frac{\mubkgsub}{\sigmavis},
\end{equation}
where \nbb represents the number of colliding bunches, $f = 11.245$~kHz is the LHC revolution frequency, and $\sigmavis$ is the detector-specific visible cross-section, which accounts for the detector efficiency and geometrical acceptance and is measured using the van der Meer scan procedure.

\begin{figure}[!htbp]
\centering
\includegraphics[width=0.8\linewidth]{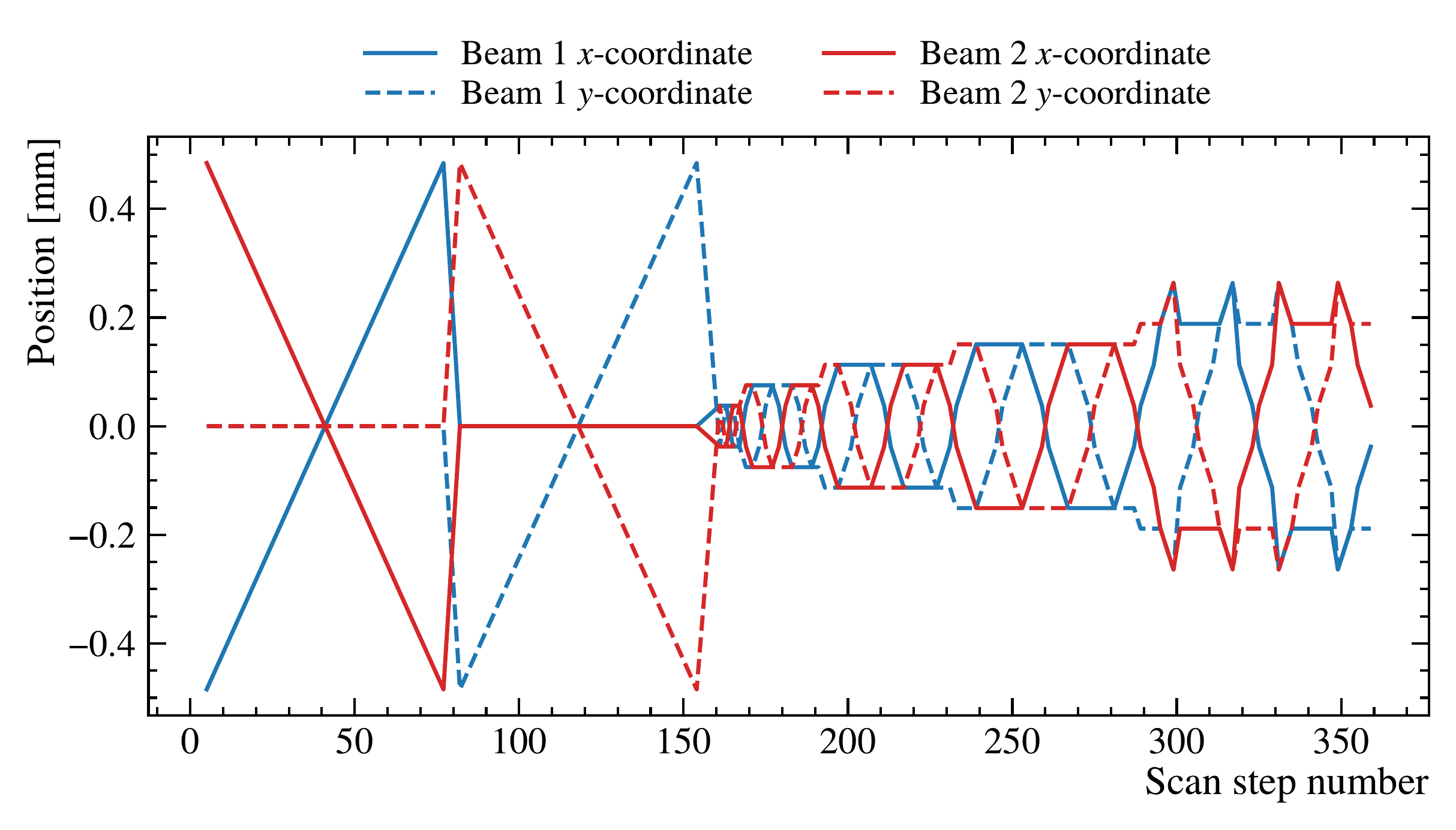}
\caption{Evolution of the LHC beams positions during a 1-dimensional sequence of a \proton\proton vdM scan, followed by a 2-dimensional sequence. A typical LHCb scan is made of three such sequences, acquired in different experimental conditions.}
\label{fig:vdm} 
\end{figure}

Any detector used to measure the instantaneous luminosity must be calibrated by measuring the visible cross-section $\sigmavis$.
The LHC performs, once per year and for each type of beam particles and centre-of-mass energy, a procedure known as the van der Meer (vdM) scan~\cite{vanderMeer}.
During this procedure, the LHC beams are first transversely offset and then brought into head-on collision before being offset again in both the $x$ and $y$ directions, using a fine offset range and performing consecutive scans in one and two dimensions.
Figure~\ref{fig:vdm} shows the sequence of a typical vdM scan.

Assuming that the beam profiles follow Gaussian distributions, the overlap integral is related to the visible cross-section by
\begin{equation}
\label{eq:vdM}
\sigmavis =
\frac{2\pi\,\mupeak\,\sigma_x\,\sigma_y}
{\langle N_1 \cdot N_2 \rangle},
\end{equation}
where $\mupeak$ is the average number of detected interactions when the beams are fully aligned, after subtracting the contributions from beam--empty and empty--beam crossings.
The beam widths in the horizontal and vertical directions are denoted by $\sigma_x$ and $\sigma_y$, respectively, and $N_{1,2}$ represent the bunch populations in beam~1 and beam~2.
Their values are obtained from the CERN TIMBER database using data from the BCTFR A6R4--B6R4 sensors~\cite{Belohrad:1267400} and are rescaled according to the total beam currents measured by the BCTDC A6R4--B6R4 detectors.

\begin{figure}[!htbp]
\centering
\includegraphics[width=0.8\linewidth]{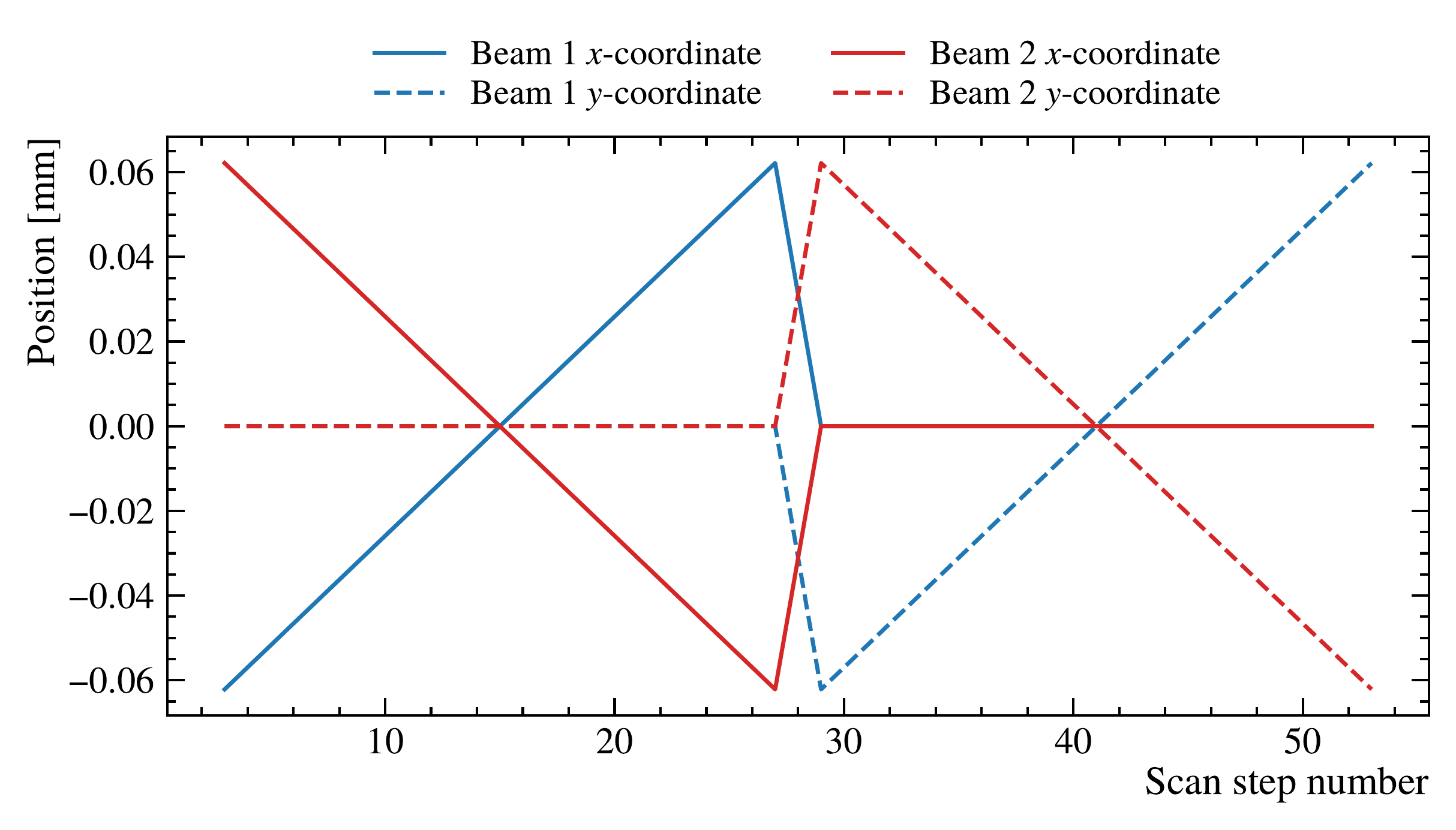}
\caption{Evolution of the LHC beams positions during a 1-dimensional \proton\proton emittance scan.}
\label{fig:emittance}
\end{figure}

Another type of beam scan, referred to as emittance scans~\cite{PhysRevAccelBeams.21.102801}, is performed routinely throughout the year to monitor the stability of the $\sigmavis$, potentially affected by changes in detector conditions.
In this case, only one-dimensional scans in $x$ and $y$ are carried out, and the beam offset range is reduced in order to minimize the required scan time and the loss of physics data.
Figure~\ref{fig:emittance} shows the sequence of a typical \proton\proton emittance scan~.

\subsection{LHC clock phase determination}
The LHC distributes a reference clock (the LHC clock) to all experiments through the Timing, Trigger and Control (TTC) system.
This clock is synchronous with the beam revolution frequency and defines the 25 ns bunch-crossing structure.
For optimal detector performance, the phase of the local detector clock must be aligned with the arrival time of particles originating from beam–beam collisions.
Any phase drift, typically caused by temperature variations, optical-fibre length changes, or electronics ageing, can lead to a mismatch between the sampling point of the front-end electronics and the true arrival time of particles.
Therefore, LHC experiments continuously monitor the LHC clock phase, defined as the timing difference between the reference clock edges and the detector signals associated with colliding bunches.
The measurement is carried out by evaluating the time offset between the collision-induced signal and the reference clock.
Maintaining this alignment is essential to avoid signal degradation, inefficiencies in time-critical detectors, and biases in luminosity or beam-monitoring measurements.

In the PLUME detector, the clock-phase measurement is performed by comparing the arrival time of PMT signals from colliding bunches with respect to the reference LHC clock distributed via the TTC system.
Since the front-end electronics are not designed to directly measure the signal arrival time with sub-nanosecond precision, the clock-phase measurement is performed using the highest-gain PMTs.
The analogue signal is split into eight replicas, each delayed by 3 ns with respect to the previous one.
This scheme allows the leading edge of the pulse to be sampled at multiple time offsets within a single bunch crossing.
The signal arrival time is then reconstructed offline  using a dedicated timing extraction procedure.
The arrival time of PMT signals from beam--beam bunch crossings is averaged over $2^{15}$ events (corresponding to a few seconds of data taking) relative to the LHC clock edge, providing a measurement of the detector clock phase with a precision of about 8 ps, as shown by variations from consecutive measurements.

Any deviation from the nominal reference time indicates a phase drift between the collision time and the LHC clock.
This information is used to adjust and synchronise the front-end sampling phase with the collision time.
Owing to the high rate and precise timing of signals from beam--beam collisions observed by PLUME, its clock-phase measurement is stable and precise, enabling the detector to contribute both to luminosity determination and to continuous timing alignment within the experiment.

\section{Detector}
\label{sec:detector}
This section provides an overview of the PLUME detector, including its layout, main components, operating principles, as well as the electronics and firmware used for readout and control.
The detector is designed to provide fast and stable signals for luminosity measurements and timing monitoring during LHC operation.

\begin{figure}[!htbp]
\centering
\includegraphics[width=0.8\linewidth]{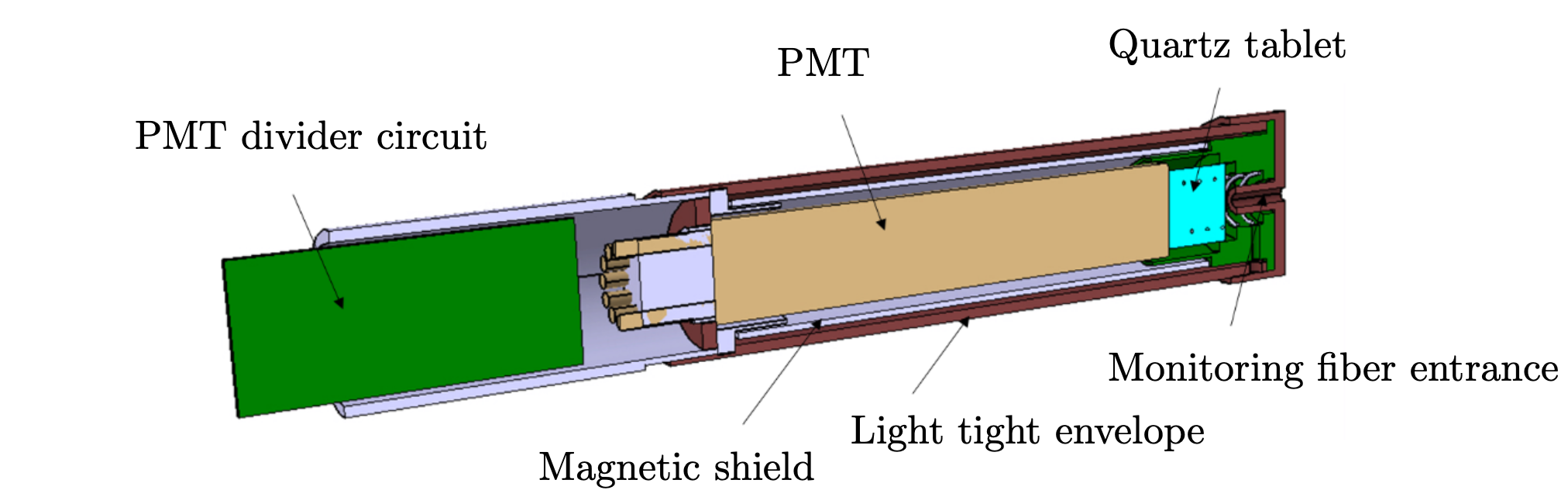}
\caption{\label{fig:PMT} Schematic view of the PLUME elementary detection module. The module is 153 mm long with a diameter of 24 mm.}
\end{figure}

Each PLUME detection unit consists of a Hamamatsu R760 head-on photomultiplier tube equipped with ten linear-focused dynodes. 
A 5 mm-thick fused silica tablet is glued to the PMT entrance window to enhance the Cherenkov-photon yield produced by incoming particles. 
Studies on simulation indicate that the addition of the fused silica tablet increases the photon yield by approximately a factor of five compared to a PMT without the tablet. 
The PMT, the fused silica tablet, and the voltage-divider circuit are enclosed within a cylindrical aluminium shield, forming the elementary detection module of the luminometer, as shown in figure~\ref{fig:PMT}.

\begin{figure}[!htbp]
\centering
\includegraphics[width=0.95\linewidth]{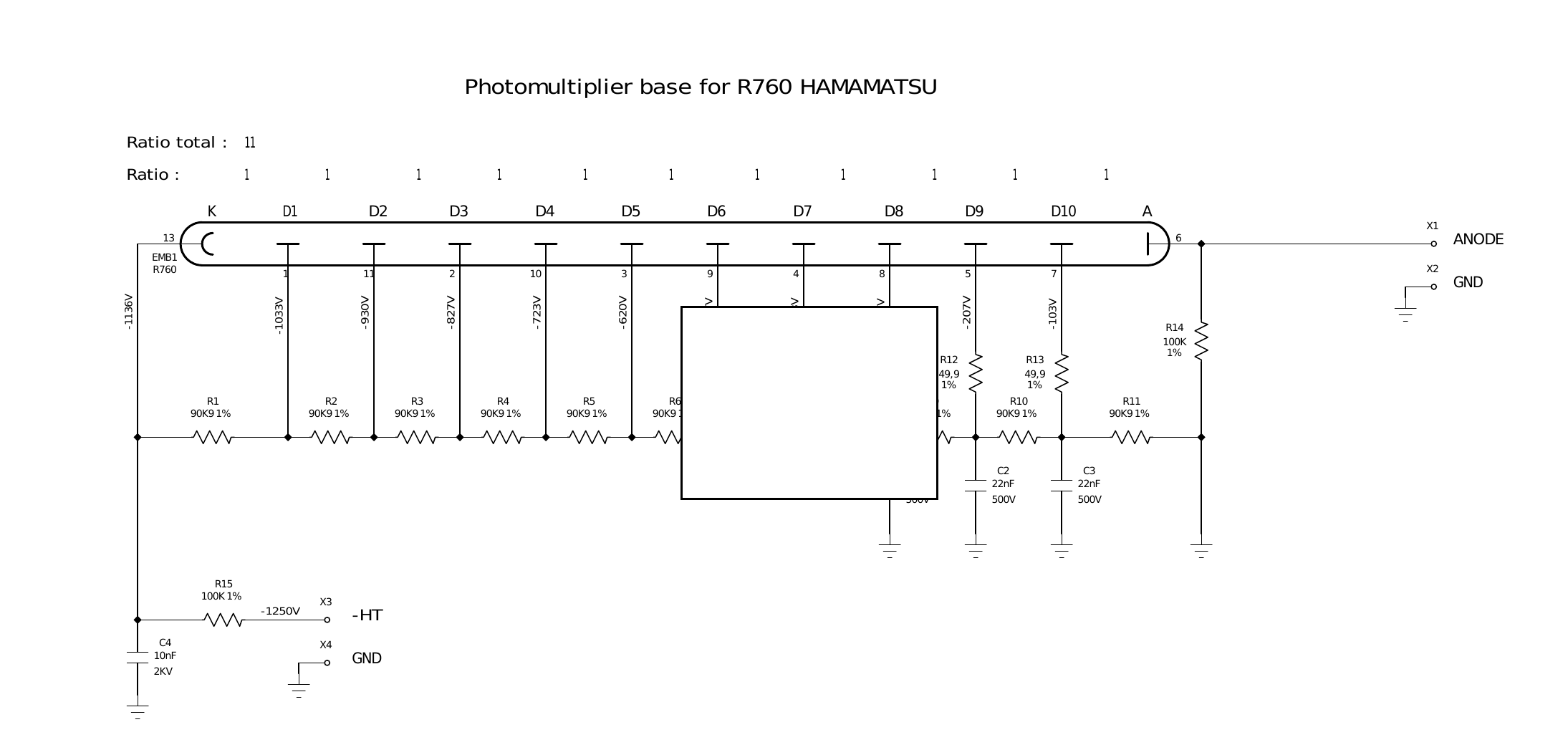}
\caption{\label{fig:voltage_divider} Schematic of the custom divider circuit used for the PLUME PMTs.}
\end{figure}

The resistive voltage divider supplied by Hamamatsu exhibits limited linearity for the typical signals observed by PLUME, particularly when accounting for the additional light produced by the fused silica tablet. 
For this reason, a new custom divider circuit has been developed, designed to deliver a maximum current of 1~mA at a bias voltage of 1000~V. 
The circuit, developed at the Laboratoire de Physique des 2 Infinis Irène Joliot-Curie (IJCLab), is shown schematically in figure~\ref{fig:voltage_divider}.

The choice of this PMT model is driven by several requirements. 
First, the signal must be fully contained within a 25~ns window to avoid spillover and ensure efficient readout. 
Second, the signal amplitude produced by a single particle must be sufficiently large to exceed the instrumental noise. 
Moreover, the harsh radiation environment requires the use of radiation-resistant materials. 
Radiation hardness is a critical constraint, as the expected dose in the relevant region is around 200~kGy after an integrated luminosity of 50~\invfb, and the 1~\mev neutron-equivalent fluence is approximately $1\times10^{14}$ neutrons/cm$^2$.
Comprehensive technical specifications of the PLUME detector are presented in the Technical Design Report in Ref.~\cite{LHCb-TDR-022}.

\subsection{Monitoring system}
\label{subsec:monitoring}
A dedicated calibration and monitoring system is implemented in PLUME to ensure the long-term stability of the detector response and the accuracy of the luminosity measurement.
The gain and efficiency of the PMTs can vary over time as a result of changes in temperature, detector occupancy, accumulated charge, and radiation damage.
Continuous monitoring is therefore required to control these effects and to apply appropriate corrections.

The PLUME monitoring system comprises a LED-based calibration system, in which controlled light pulses are periodically injected into each PMT via quartz fibres during suitable gaps in the LHC filling scheme.
The PMT response to the injected light is monitored at regular intervals, while the stability of the LED light output is independently tracked using PIN photodiodes located close to the LEDs in a region of reduced radiation exposure.
In addition, radiation-induced degradation of the quartz fibres is monitored using dedicated fibres looped back to reference PMTs.

The monitoring system was originally designed to provide gain stabilisation of the PMTs during operation.
Results obtained during data taking indicate that this approach allows the PMT gain to be controlled with a precision of a few percent. 
The achievable precision is mainly limited by the accuracy with which the degradation of the optical fibres can be measured.
To further improve the gain calibration accuracy, a complementary and data-driven technique has been developed and is described in Sec.~\ref{subsec:gain_stability}.

The monitoring system proved to be valuable beyond its original purpose during the commissioning of the PLUME backend boards firmware. 
In particular, the ability to send LED pulses to individual PMTs and in correspondence of a given bunch-crossing was instrumental both for debugging the firmware and for characterizing the detector response.

The data collected by the monitoring system are fully integrated into the PLUME readout and control infrastructure and are available for both online and offline analysis.

\subsection{Electronics and readout}
\label{subsec:FE_BE_firmware}
PLUME data are made available both online, for low-latency luminosity measurements, and offline, together with the rest of the LHCb dataset. 
Both functionalities are handled by the same readout system, which interfaces with the LHCb Experiment Control System (ECS) and Data Acquisition (DAQ) system.
The electronics chain, which largely reuses components developed for the LHCb calorimeters, consists of a front-end (FE) board, the FEB~\cite{LHCb-DP-2022-002}, and a back-end (BE), the PCIe40 board~\cite{TELL40}.
The PLUME readout elements are organised in a dedicated partition within the LHCb readout framework for ECS and DAQ.

\subsection{Front-end electronics}
The FE boards used to read out the PMTs are identical to those designed for the LHCb Upgrade I calorimeter readout.
Their role is to amplify, shape, and integrate the signals from the detector PMTs using a custom four-channel ASIC, ICECAL~\cite{Picatoste_2012}, developed for the ECAL readout.
Each FE board provides 32 input channels through eight ICECAL ASICs.
PLUME employs five FE boards: two read out the 44 PMTs used for luminosity measurements (22 PMTs per board), one is dedicated to the monitoring system, and two read out the four PMTs used for beam-phase timing.
For the timing readout, each PMT is connected to 16 channels: eight sampling the signal in the nominal bunch crossing and eight storing the signal from the preceding bunch crossing.
The processed signals are digitised by 12-bits ADCs, and the resulting digital data are handled by FPGAs.
After formatting, the data are transmitted to the back-end electronics via optical fibres.
Each PMT signal is encoded in 12 bits (4096 ADC counts), with each count corresponding to a charge of 11~fC.
This dynamic range is sufficient for both proton--proton and ion runs.

All FE boards are housed in a crate located in the LHCb cavern, which supplies the required voltages.
Within the same crate, a control board (3CU)~\cite{LHCb-DP-2022-002} distributes the clock, commands, and configuration from the global LHCb systems to all FE boards housed in the crate.
For the PLUME setup, a single 3CU board is required, employing the same model as used for the calorimeter system.

\subsection{Back-end eletronics}
The back-end electronics is implemented in the common LHCb readout board for the LHCb upgrade, the PCIe40 board, which is a PCIe based electronics boards hosting an Intel Arria 10 FPGA.
The boards dedicated to data processing are called TELL40.
Two TELL40 boards are assigned to PLUME: the first one is used to readout the 44 PMTs dedicated to the luminosity measurement, while the second one is used for the beam-phase timing PMTs and the monitoring system.

\subsubsection*{Firmware for online luminosity determination}
\begin{figure}[!htbp]
    \centering
    \includegraphics[width=0.95\linewidth]{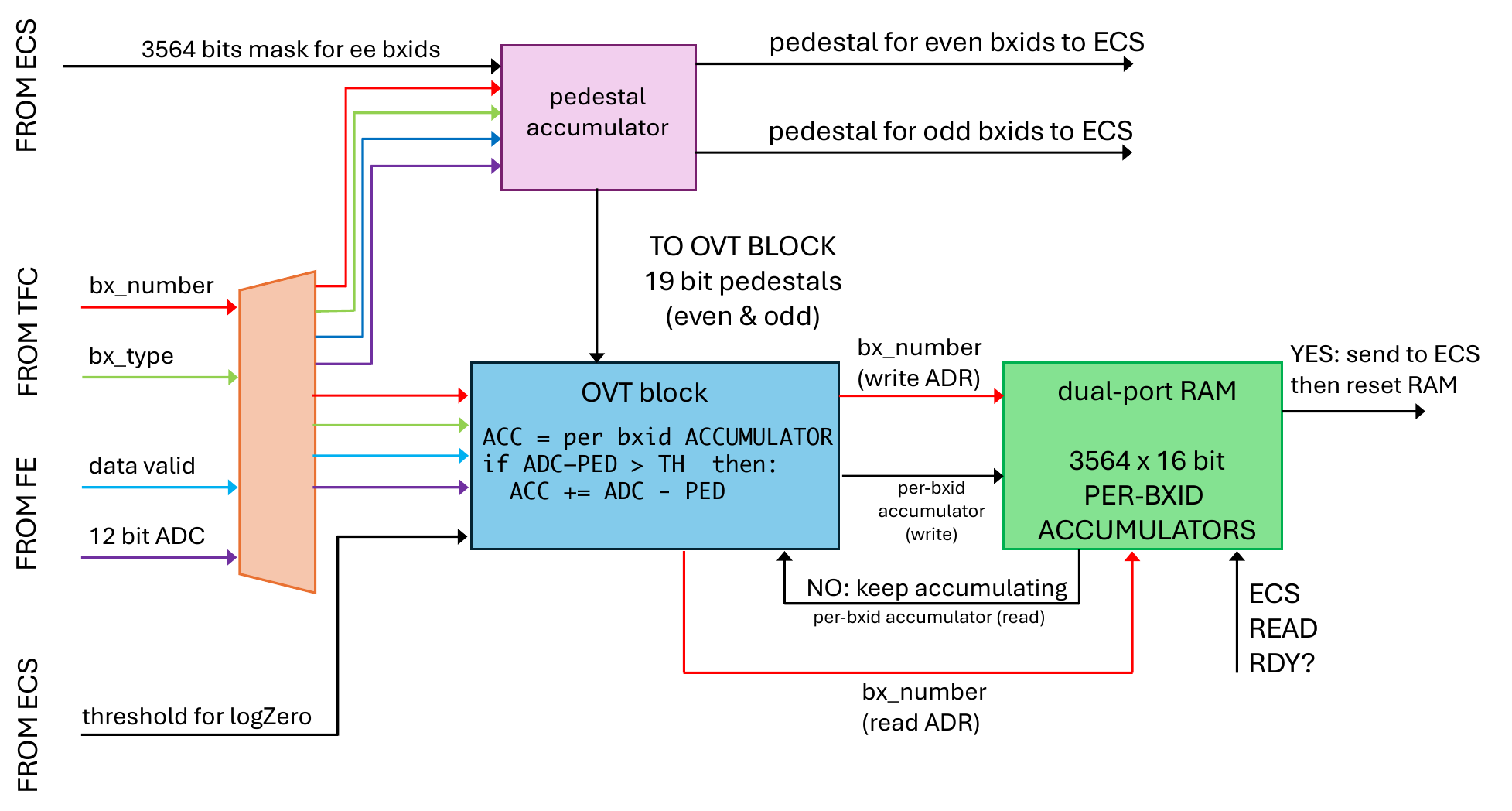}
    \caption{Sketch representing the main logic blocks used to calculate online luminosity for each PMT in the TELL40 firmware.}
    \label{fig:tell40_fw_lumi}
\end{figure}

The firmware developed for the online luminosity determination is represented schematically in figure~\ref{fig:tell40_fw_lumi}.
The TELL40 receives the 12-bit ADC counts from the FE boards, aligned with the LHC bunch-crossing (BX) number and type provided by the experiment’s Timing and Fast Control (TFC) signal.
This information is used to compute pedestals for each PMT, separately for even and odd bunches.
This separation is necessary because the FE boards are equipped with two charge integrators per channel, one for even and one for odd bunch crossings, which may exhibit different baseline values.
The pedestal is calculated as the average over approximately one million $(2^{20})$ events, with a precision of two decimal digits.
Depending on the LHC filling scheme, this corresponds to new pedestal values becoming available roughly once per second.
These values are sent both to the ECS registers for offline monitoring and to the next logic block in the TELL40, known as the \texttt{OVT} block.

The \texttt{OVT} block receives the same inputs as the pedestal accumulator, together with a threshold value configurable via the ECS.
The pedestal is subtracted from the 12-bit ADC value coming from the FE board, and the resulting signal is compared with this threshold.
If the value exceeds the threshold, a counter is incremented by one.
This counter is stored in a dual-port RAM comprising 3564 16-bit slots, with the bunch-crossing number used as the RAM address.
The dual-port RAM also receives a \texttt{READ READY} signal from the ECS.
When this signal is inactive, the \texttt{OVT} block continues accumulating counts by retrieving the existing per-BX 16-bit value from RAM as soon as the corresponding address is issued.
When a valid \texttt{READ READY} signal is received, the accumulation stops and the contents of the dual-port RAM are transferred to the ECS registers.
These registers are then read immediately by a system task running on the server hosting the TELL40.
This task polls the TELL40 every 2.4 seconds, which corresponds to the total time required to transmit the luminosity information to the LHC.
Of this interval, 2.0 seconds are configurable within the TELL40 and correspond to the time allocated for statistical accumulation, while the remaining 0.4 seconds account for ECS processing.
This latency is well within the LHC requirement for luminosity-leveling feedback, which must remain below 3.0 seconds.
The system’s inactive time, occurring during the readout phase, is on average approximately 50 ms, a negligible contribution compared with the total 2.4 seconds. 

\subsubsection*{Firmware for beam-phase monitoring}
\begin{figure}[!htbp]
    \centering
    \includegraphics[width=0.95\linewidth]{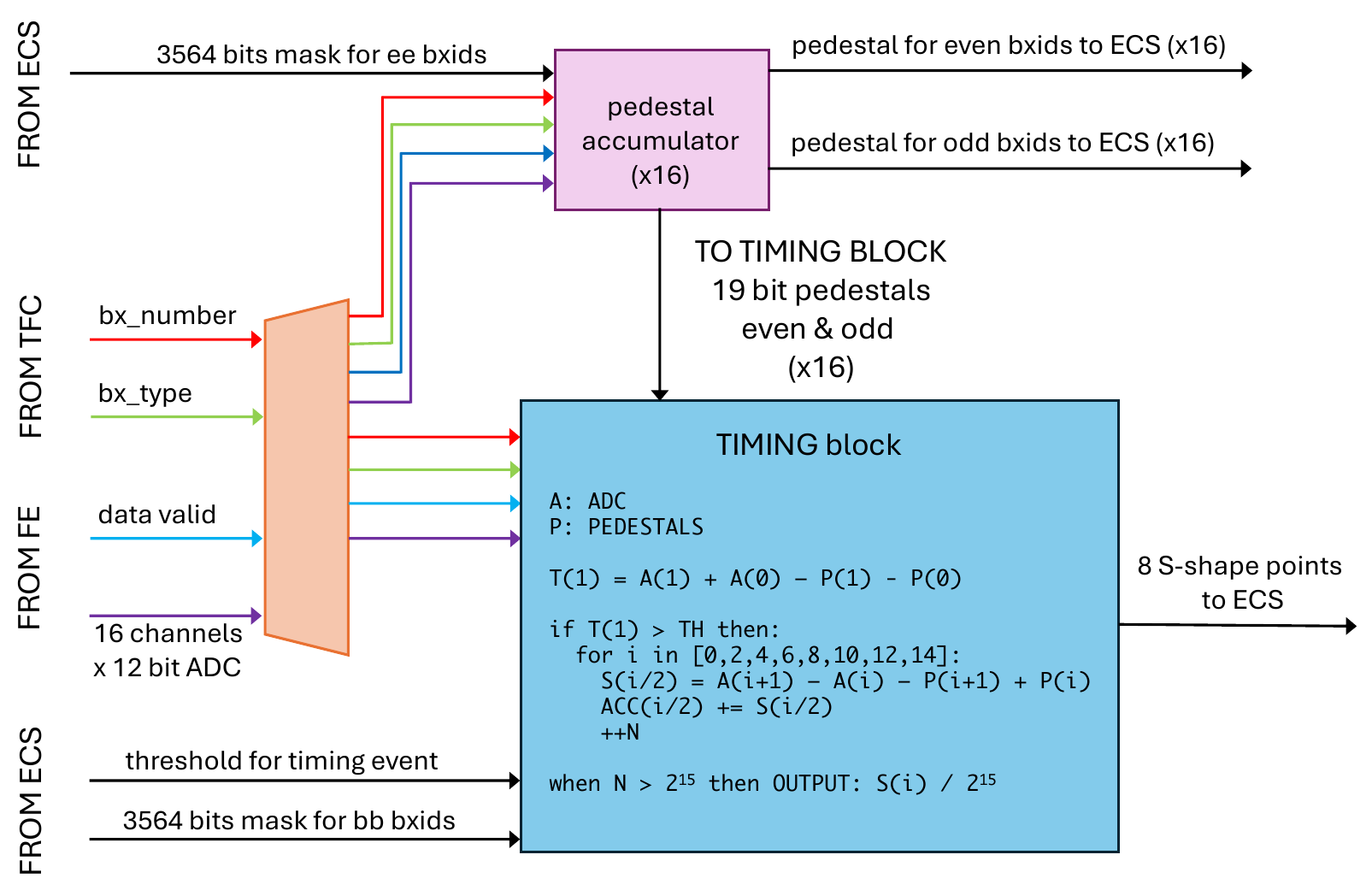}
    \caption{Sketch representing the main logic blocks used to calculate beam-phase timing for each timing PMT in the TELL40 firmware. The $i$ index stands for the FE board channels and can assume all the even values between 0 and 15.}
    \label{fig:tell40_fw_timing}
\end{figure}

The firmware developed for the beam-phase determination is represented schematically in figure~\ref{fig:tell40_fw_timing}.
The logic used for pedestal computation is identical to that of the baseline firmware previously described. 
The distinctive feature that needs to be taken into account by this version is that the signal from the four timing PMTs are split into eight identical copies, with each copy delayed by an increasing amount and routed to a different FE board channel.
An additional eight channels of the same front-end board are used to store information from the bunch crossing preceding the nominal one, resulting in a total sampling window of 50 ns for each PMT and allowing the signal integral to be gradually distributed across two consecutive bunch crossings.
By taking the difference between pairs of front-end channels that sample the same PMT signal with different delays, one corresponding to the nominal bunch crossing and the other to the preceding one, a smooth transition curve is reconstructed.
This curve represents the cumulative response of the signal waveform as a function of time.
To address this, a dedicated \texttt{TIMING} block has been developed for each of the timing PMTs.
Following the selection of beam--beam bunch crossings, a charge-based selection is applied to retain only signals with sufficiently large amplitude, which
provide an improved signal-to-noise ratio and time resolution.
This selection is implemented by exploiting the integrated charge information available from consecutive front-end samples.

For the selected events, the digitised samples are pedestal-subtracted and combined to reconstruct the cumulative timing response of the PMT signal.
The resulting distributions are accumulated over $2^{15}$ events and periodically transmitted to the Experiment Control System (ECS).

Custom scripts can read these values and fit the points with an error function to determine the inflection point, which is then used as the event timestamp.
It should be noted that this is an average timestamp, not a per-event measurement.
However, offline toy studies have shown that the average of many per-event timestamps is equivalent to the computed average timestamp.
This approach is sufficient to provide a beam-phase timing measurement approximately every five seconds.

\section{Performance}
\label{sec:performance}
This section presents a detailed account of the commissioning and operation of PLUME during the 2024–2026 data-taking periods.
It describes the activities performed to bring the system into operation, the commissioning procedures, the achieved performance, and the adjustments and optimisations applied during routine operation.

\subsection{Gain stability}
\label{subsec:gain_stability}
Ensuring detector stability throughout the data-taking period is essential.
Any uncorrected variation in PMT gain can compromise the vdM calibration. Therefore, stable operating conditions for the PMTs must be maintained.
To this end, a dedicated procedure has been implemented.

\begin{figure}[!htbp]
    \centering
    \includegraphics[width=0.38\linewidth]{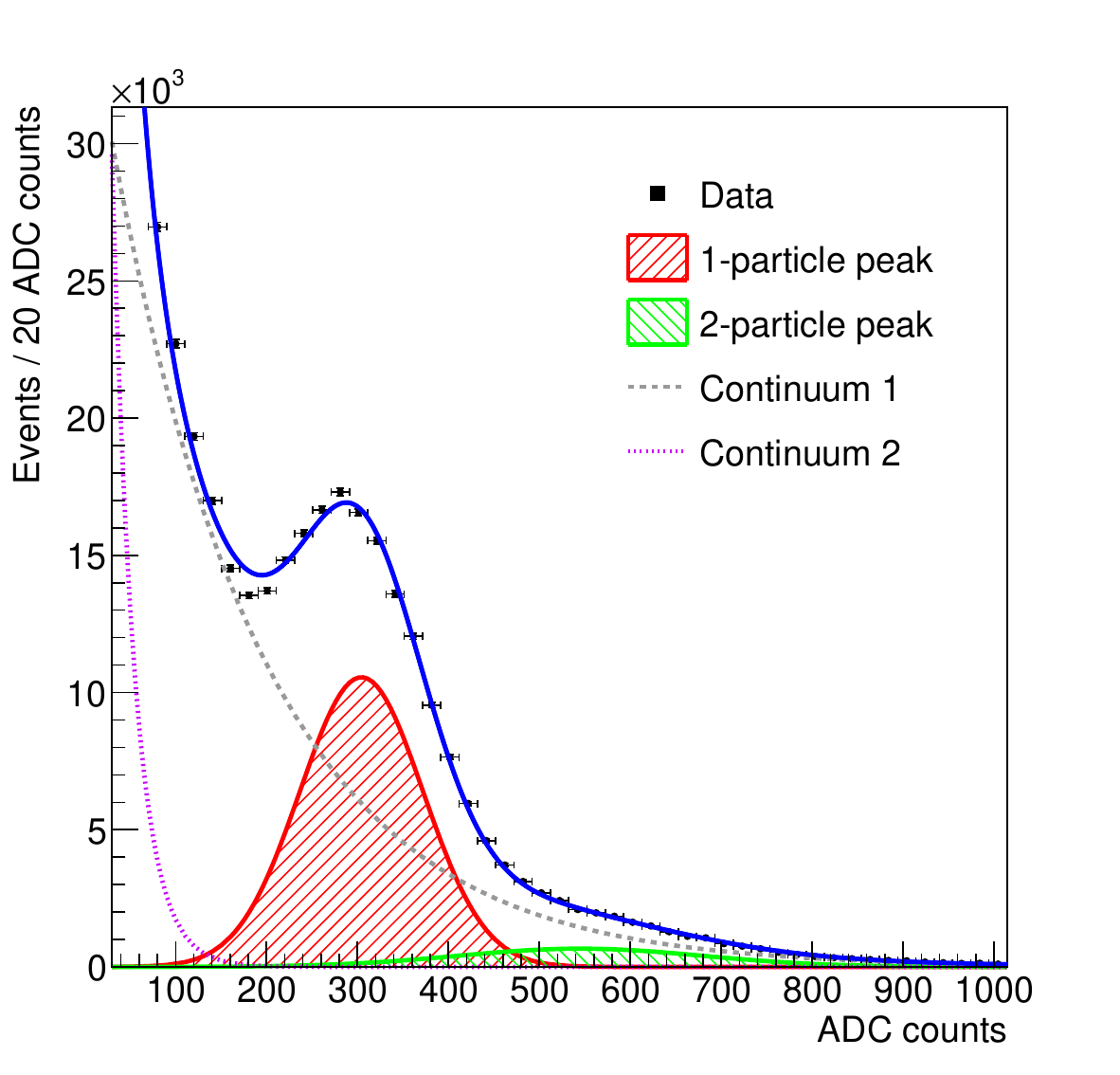} 
    \includegraphics[width=0.45\linewidth]{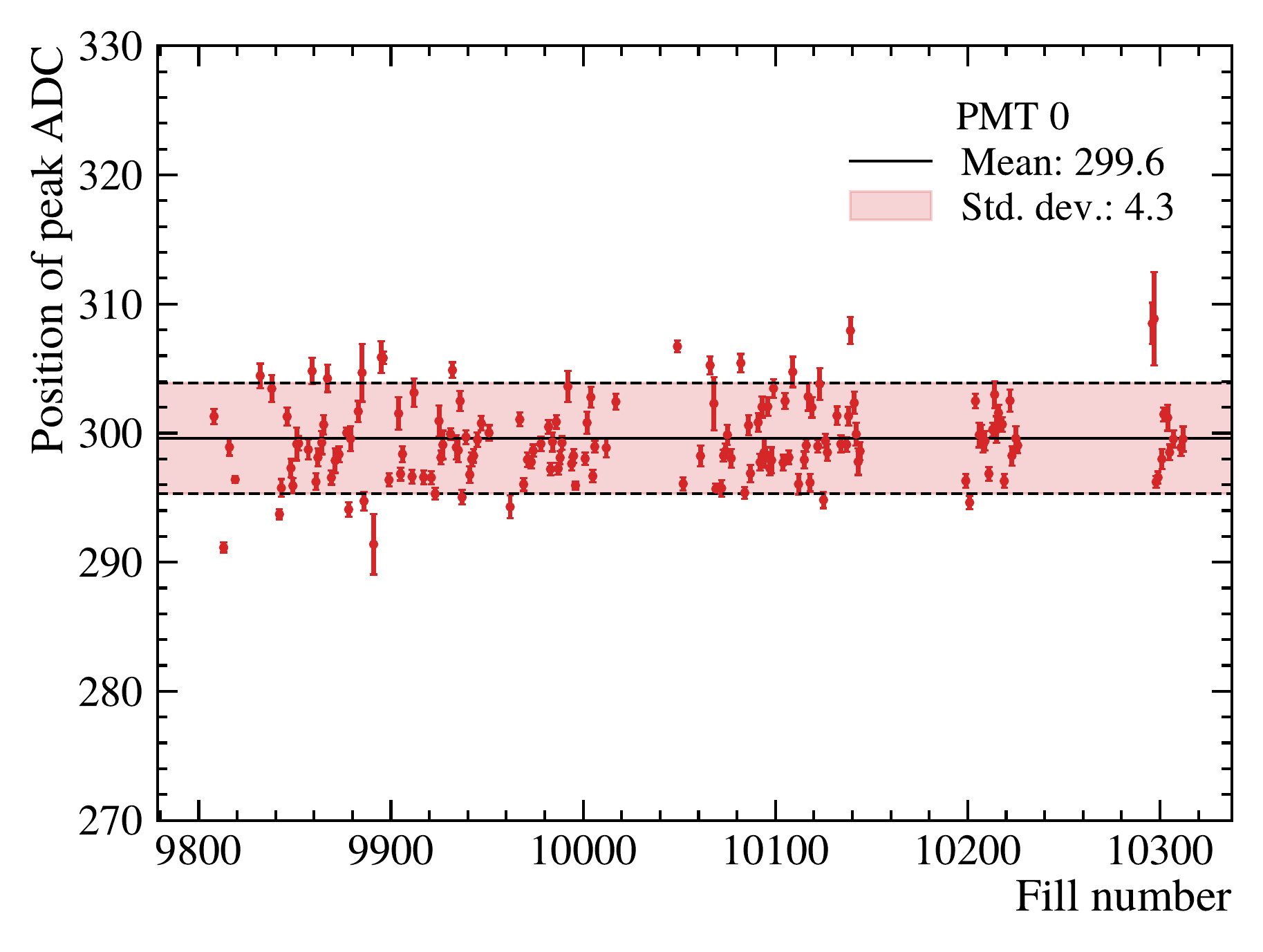}
    \includegraphics[width=0.45\linewidth]{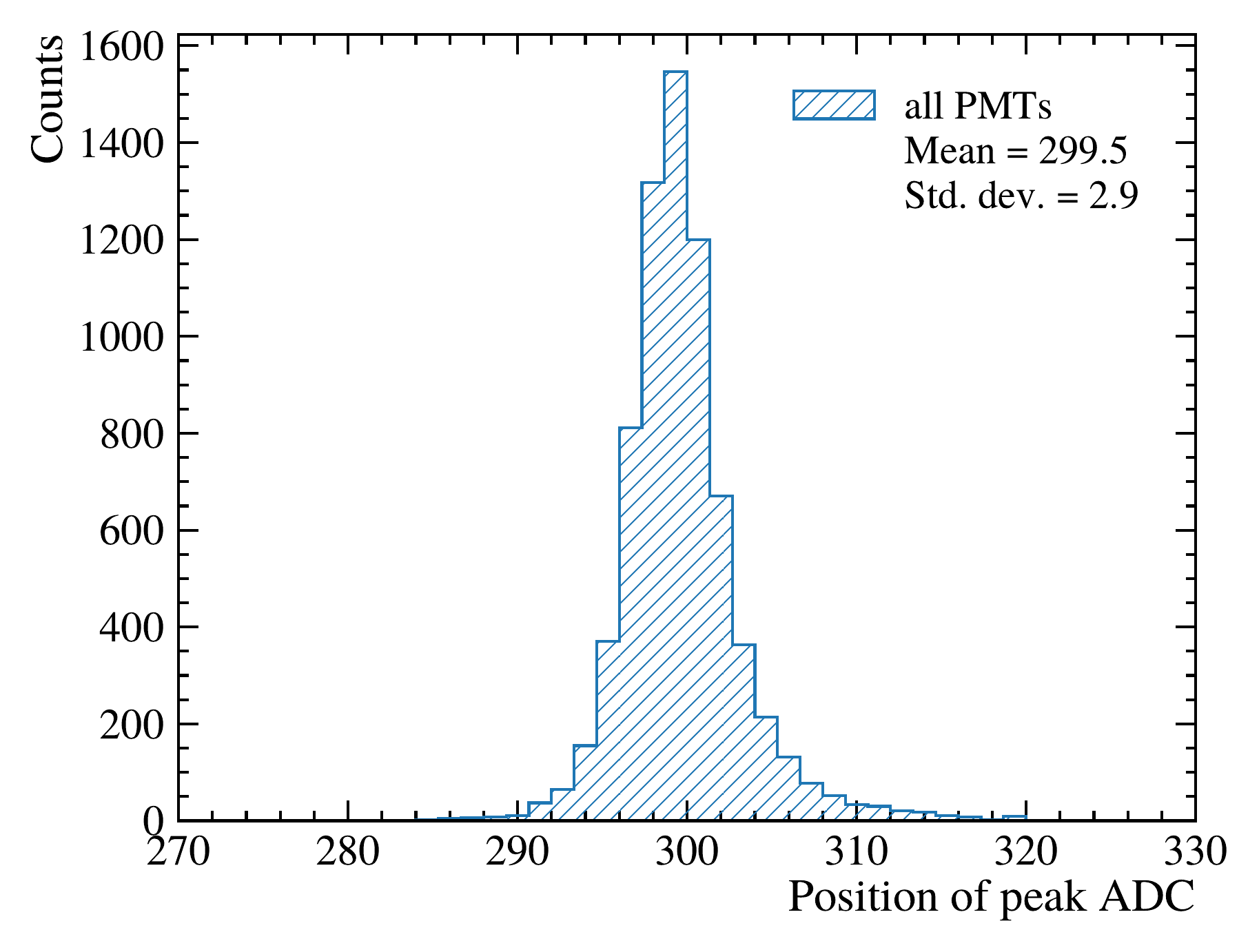}
    \caption{Top left: ADC spectrum obtained from the LHCb monitoring system for PMT~2 during LHC fill 9931, with the fit results overlaid. Top right: Mean value of the charge distribution associated with quasi-perpendicular particle incidence on the PMT window for PMT~0, shown as a function of the LHC fill number. Bottom: Distribution of the mean charge values corresponding to quasi-perpendicular particle incidence on the PMT fused silica windows, accumulated over the entire 2024 $pp$ data-taking period. Data from all PMTs are aggregated.}
    \label{fig:adc_monitoring}
\end{figure}

After the beams are dumped at the end of each physics fill, the ADC spectra accumulated during the fill by the LHCb monitoring system are collected for all PMTs and fitted.
The fit model includes a Gaussian component describing the charge distribution associated with particles impinging quasi-perpendicularly on the PMT window, together with a second Gaussian component accounting for events in which two particles enter the PMT within the same bunch crossing.
Additional exponential components are included to model events corresponding to particles incident on the PMT window at angles significantly different from normal incidence.
An example of such a fit is shown in the top-left panel of figure~\ref{fig:adc_monitoring}.

To monitor gain stability, the fit extracts the mean value of the charge distribution associated with quasi-perpendicular particle incidence on the PMT window, corresponding to the red peak in the top-left panel of figure~\ref{fig:adc_monitoring}.
For a nominal gain of $1.5 \times 10^{5}$, this mean charge value is expected to
be centred at approximately 300 ADC counts.
The top-right panel of figure~\ref{fig:adc_monitoring} shows the evolution of this
mean charge value as a function of the LHC fill number for PMT 0 over the course of 2024, taken as a representative example.
The average relative deviation from the target value is of the order of 1\%.
The bottom panel displays the distribution of the mean charge values associated
with quasi-perpendicular particle incidence on the PMT fused silica windows,
accumulated over the entire 2024 $pp$ data-taking period for all PLUME PMTs. A similar performance has been achieved for 2025 and 2026 data-taking periods.

Any shift in the extracted mean charge value, corresponding to a variation in
PMT gain, directly translates into a bias in the instantaneous luminosity
measurement.
Since the gain calibration is performed independently for each PMT, toy Monte
Carlo simulations have been used to estimate the resulting uncertainty on the
PLUME average instantaneous luminosity.
In these simulations, the ADC histograms for each PMT collected by the LHCb
monitoring system are randomly shifted by an amount sampled from a Gaussian
distribution with zero mean and a width equal to the observed spread of the mean
ADC charge values relative to the target.
The \emph{logZero} method is then used to compute the average \mubkgsub for a
given run number, which is compared to the non-shifted case.
This procedure is repeated multiple times, and the spread of the resulting
\mubkgsub values is taken as the uncertainty associated with imperfect gain
calibration over the year.
The total effect on \mubkgsub, averaged over all PMTs, is found to be approximately 0.4\%.

\begin{figure}[!htbp]
\centering
\includegraphics[width=0.49\linewidth]{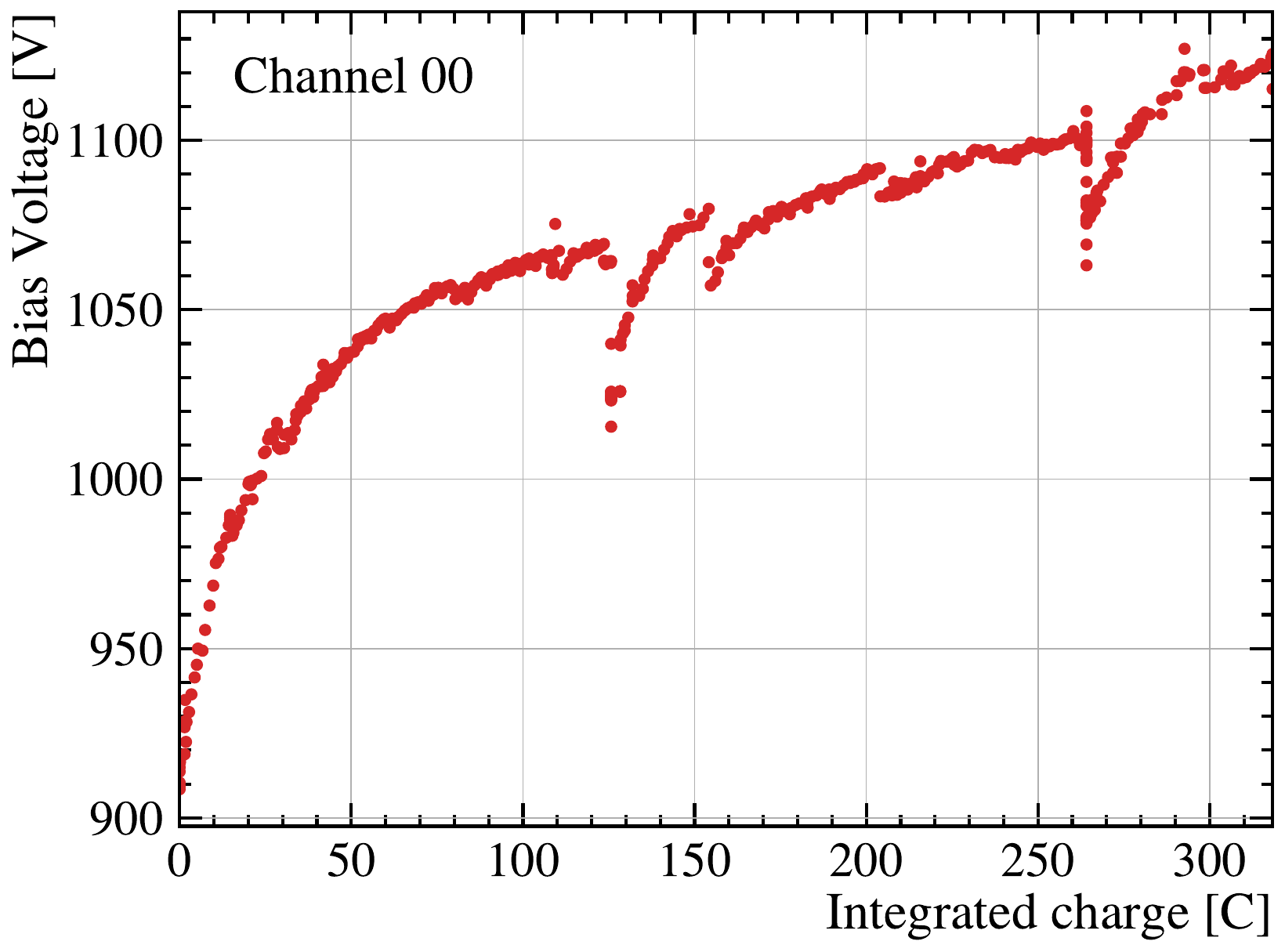}
\includegraphics[width=0.49\linewidth]{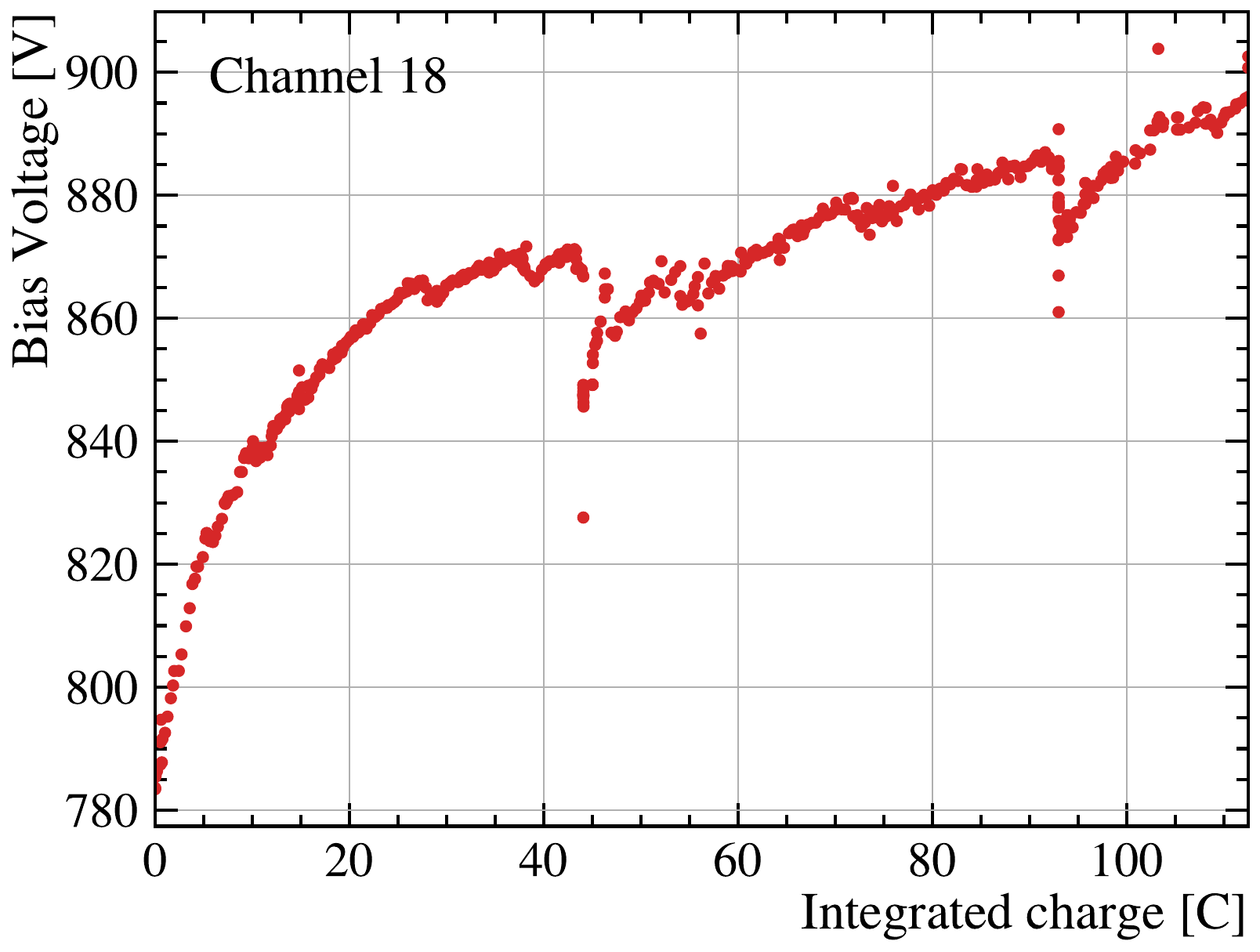}
\caption{Bias high voltage applied to (left) PMT~0 and (right) PMT~18 to maintain a constant gain as a function of the total charge collected by the corresponding PMT.
The high-voltage correction is computed at the end of each LHC fill using the gain-adjustment procedure.
The apparent clustering of several corrections at the same integrated charge corresponds to the ramp-up period after the LHC Year-End Technical Stop.
During this period, the integrated charge accumulated by the PMTs is negligible compared to that collected during physics operation, causing multiple corrections to appear at nearly the same integrated charge.}
\label{fig:hv_vs_charge}
\end{figure}

The increase in the PMT high-voltage supply, resulting from the gain adjustment procedure, is continuously monitored to determine if any action is required during one of the LHC end-of-the year technical stops, such as replacing PMTs that reach the maximum allowed voltage (1375 V).
Figure~\ref{fig:hv_vs_charge} shows the high voltage required to maintain a constant PMT gain as a function of the integrated charge collected by PMT~0 and PMT~18 during the 2024--2026 data-taking periods.
These two channels are shown as representative examples of the lower and upper bounds of PMT ageing within the PLUME detector, with PMT~0, located closest to the beam pipe, accumulating the largest integrated charge, and PMT~18, positioned farther away, collecting the smallest.
A simple extrapolation of the observed trend for the PMT~0 indicates that bias voltages as high as 1375~V would only be reached for integrated charges of the order of 1~kC or more, corresponding to about 100~\invfb integrated delivered luminosity, well beyond the expected exposure by the end of Run 4.

Starting from 2026, an alternative gain-calibration procedure was developed and deployed. In the original approach described above, the calibration relies on ADC spectra histograms produced by the LHCb monitoring system for each PMT. 
However, the statistics accumulated in these histograms can be insufficient to perform accurate fits, particularly during short fills or data-taking periods characterized by a low average interaction rate, such as PbPb collisions.

To overcome this limitation, a new calibration strategy was introduced based on the direct use of the TELL40 RAM memories as histograms, with a bin width of 8 ADC counts. 
Access to the ADC information within the TELL40 allows the full LHC bunch-crossing rate to be exploited, avoiding the statistical limitations of the LHCb monitoring system, which samples the collision rate at only a few kHz.

\begin{figure}[!htbp]
\centering
\includegraphics[width=0.42\linewidth]{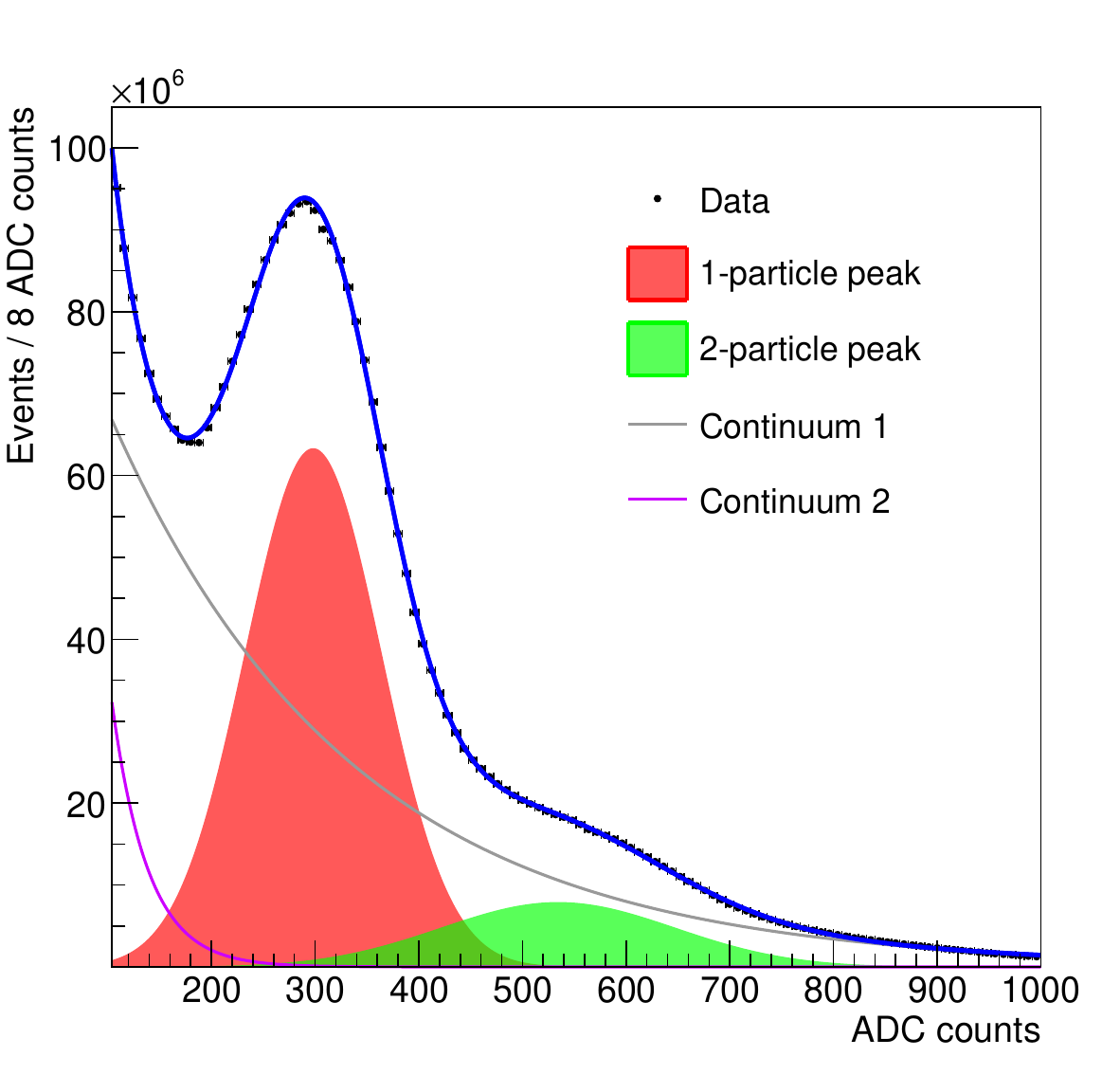} 
\includegraphics[width=0.57\linewidth]{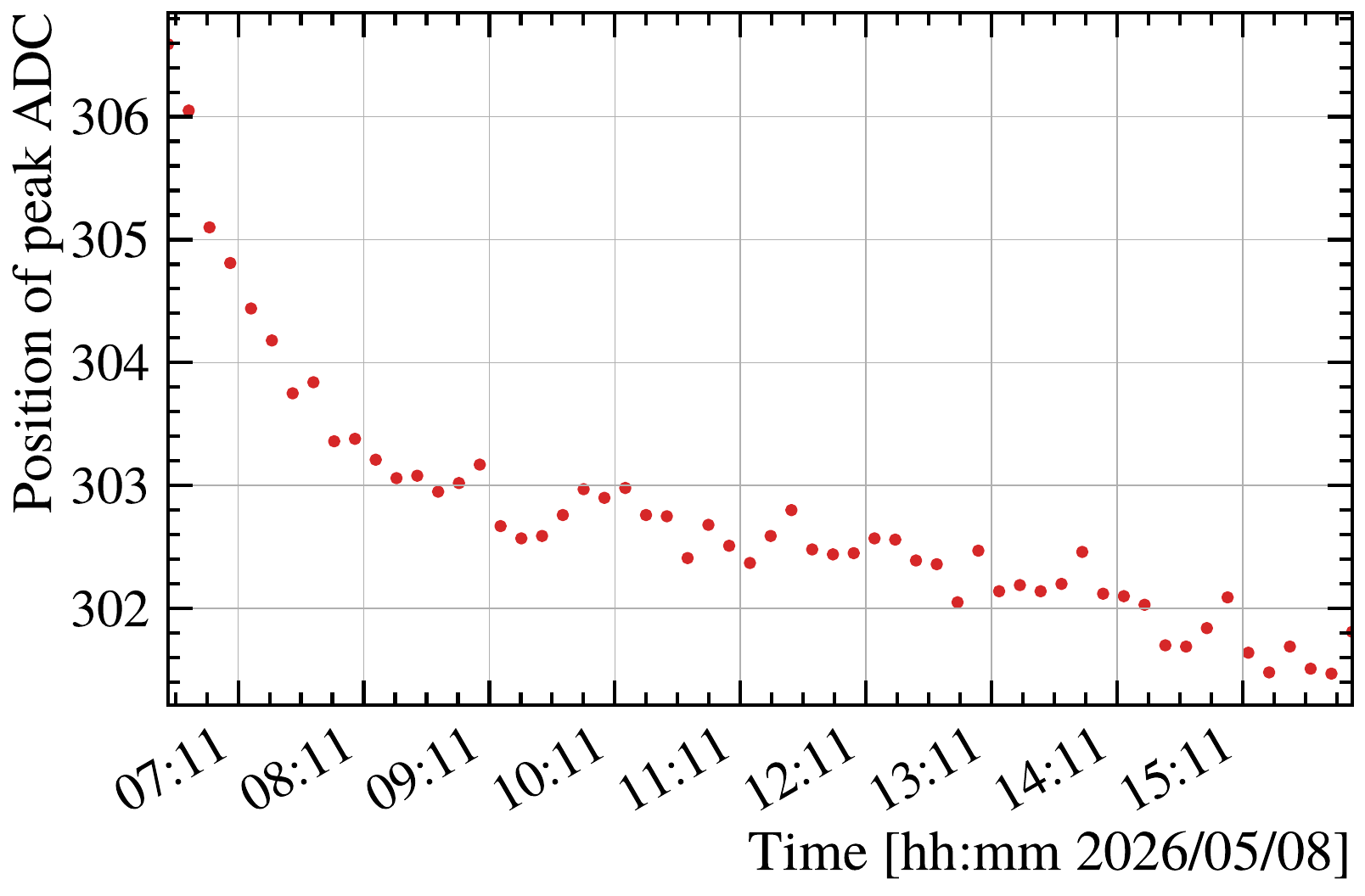} 
\caption{Left: ADC distribution measured for PMT 0 using 10 minutes of data collected by the TELL40 firmware during a nominal $pp$ fill. Owing to the full bunch-crossing rate available from the TELL40 readout, the histogram contains significantly more entries than those produced by the LHCb monitoring system, such as the example shown in the upper-left panel of figure~\ref{fig:adc_monitoring}. Right: Evolution of the mean charge value extracted from the fit to the ADC distribution during a representative fill.}
\label{fig:TELL40_adc_monitoring}
\end{figure}

Thanks to the significantly larger available statistics, the TELL40 RAMs can be read out periodically during a fill, eliminating the need to accumulate statistics over the entire fill, as required in the original approach based on monitoring histograms. 
This enables a more precise and more frequent determination of the mean charge corresponding to quasi-perpendicular particle incidence on the PMT window. 
In the implemented scheme, the TELL40 RAMs were read out every 10 minutes, enabling frequent monitoring of the PMT gain during a fill. 
The procedure was used to assess the stability of the gain calibration and to study the feasibility of more frequent high-voltage adjustments in future operations. 
An example of a histogram acquired during a representative $pp$ fill is shown in figure~\ref{fig:TELL40_adc_monitoring}, together with the evolution of the mean charge value throughout the fill.

\subsection{Luminosity determination}
\label{subsec:lumidete}
The luminosity determination in PLUME is based on the measurement of the average interaction rate and on the calibration of the corresponding visible cross-section.
The general formalism relating these quantities has been introduced in the previous section (see Sec.~\ref{sec:intro}).
Here, the focus is placed on the practical implementation of the luminosity measurement, including the determination of the interaction rate from PLUME data and the calibration of the visible cross-section using the van der Meer procedure.

\subsubsection*{Interaction rate determination}

\label{subsubsec:mu_determination}
The average number of interactions is determined for each PMT and bunch crossing by the TELL40 firmware using the first-order expression derived from Eq.~\eqref{eq:mu_def}, neglecting the second-order corrections, which are below $10^{-3}$.

\begin{figure}[!htbp]
\centering
\includegraphics[width=0.51\linewidth]{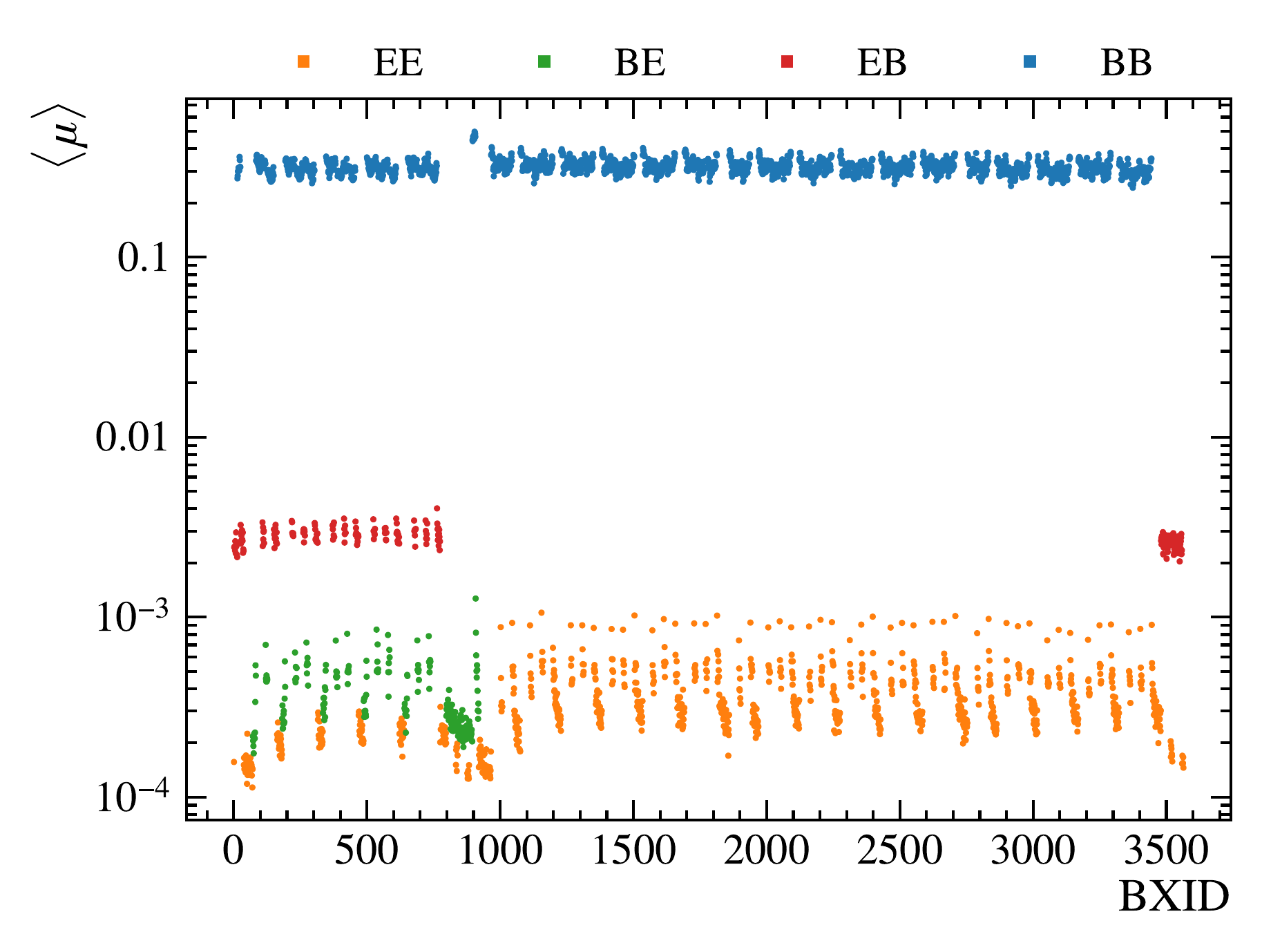} 
\includegraphics[width=0.48\linewidth]{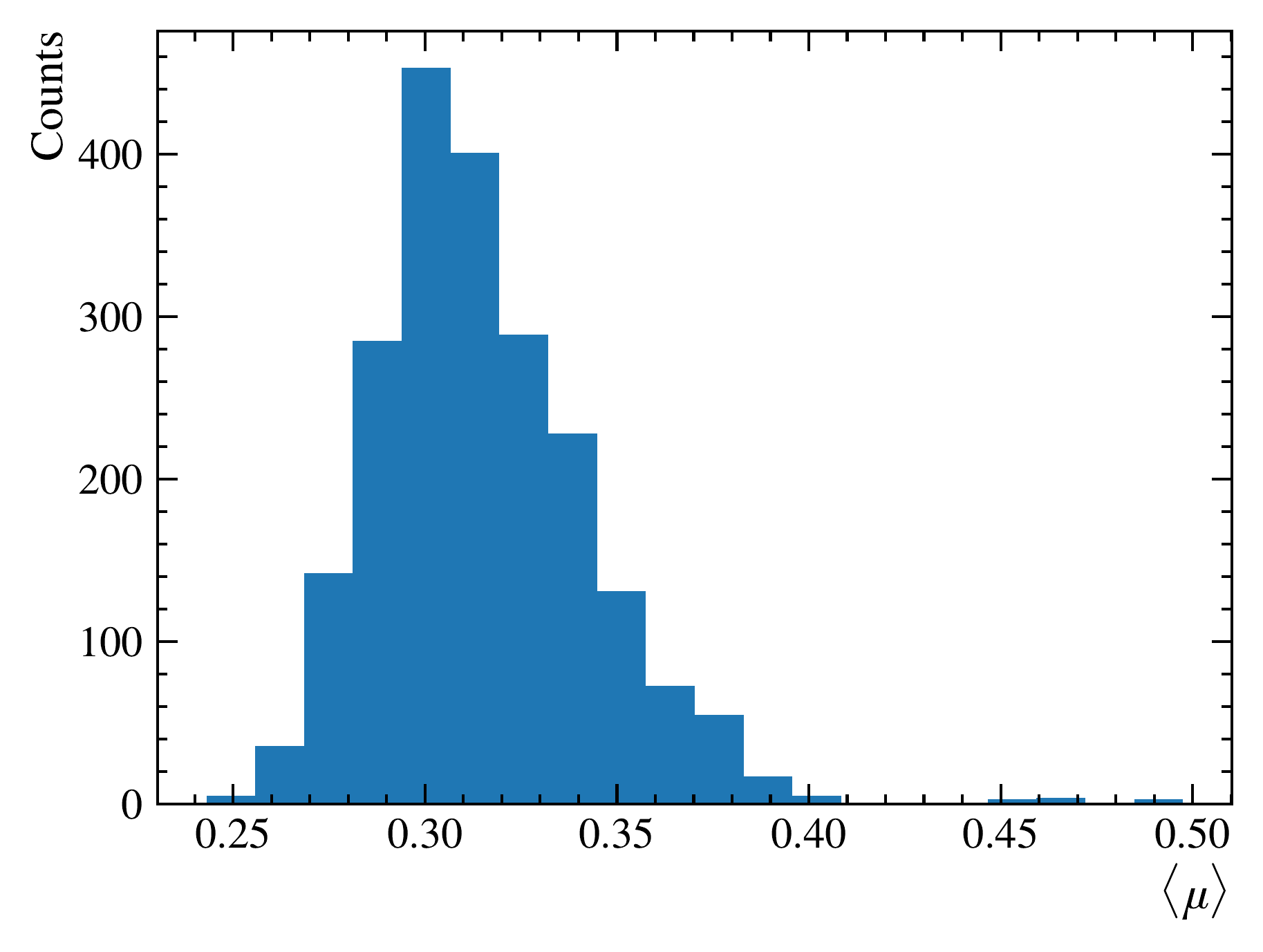} 
\caption{Left: Visible interaction rate averaged over all PLUME PMTs, $\langle \mu \rangle$, as a function of the BXID for the different bunch-crossing types. Right: Distribution of $\langle \mu \rangle$ for colliding bunch-crossings.}
\label{fig:muvis_bxid}
\end{figure}

Once this information is read out, a system task computes the average $\mu$ for each bunch-crossing type and evaluates \mubkgsub for every PMT, according to Eq.~\ref{eq:mu_bkg_sub}. From this quantity and the corresponding visible cross-sections, 44 independent luminosity values are calculated and subsequently averaged to provide the final luminosity measurement to the LHC. 
The visible interaction rate, $\mu$, is also averaged over the 44 PMTs and it is shown in figure~\ref{fig:muvis_bxid}, together with the corresponding distribution restricted to beam--beam bunch crossings.

\subsubsection*{Calibration with van der Meer and emittance scans}
\label{subsec:vdM}
\begin{figure}
    \centering
    \includegraphics[width=\linewidth]{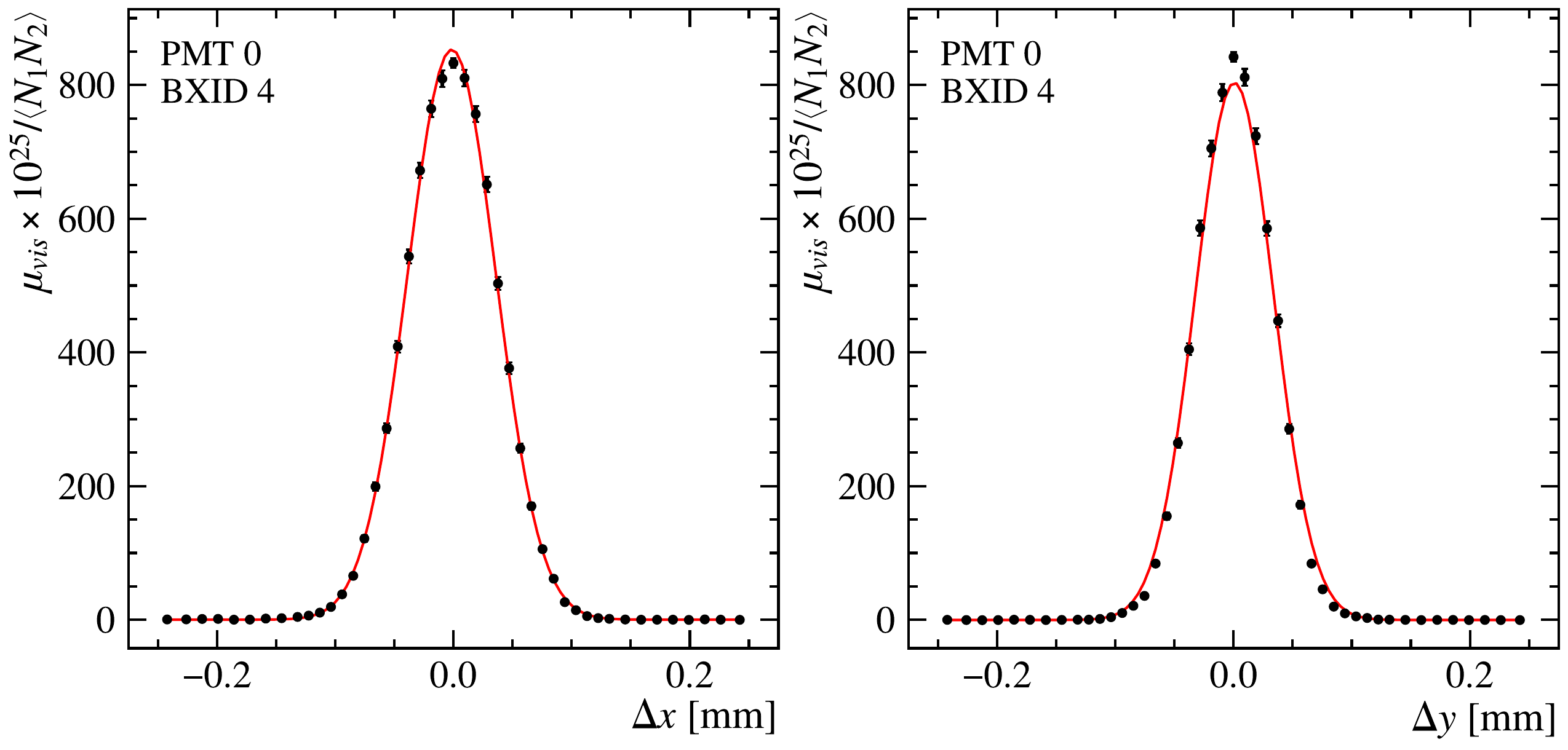}
    \caption{Fits to a 1-dimensional scan of a $p{\rm O}$ vdM sequence, LHC fill 10782.}
    \label{fig:pO_scan_fits}
\end{figure}

\begin{figure}
    \centering
    \includegraphics[width=\linewidth]{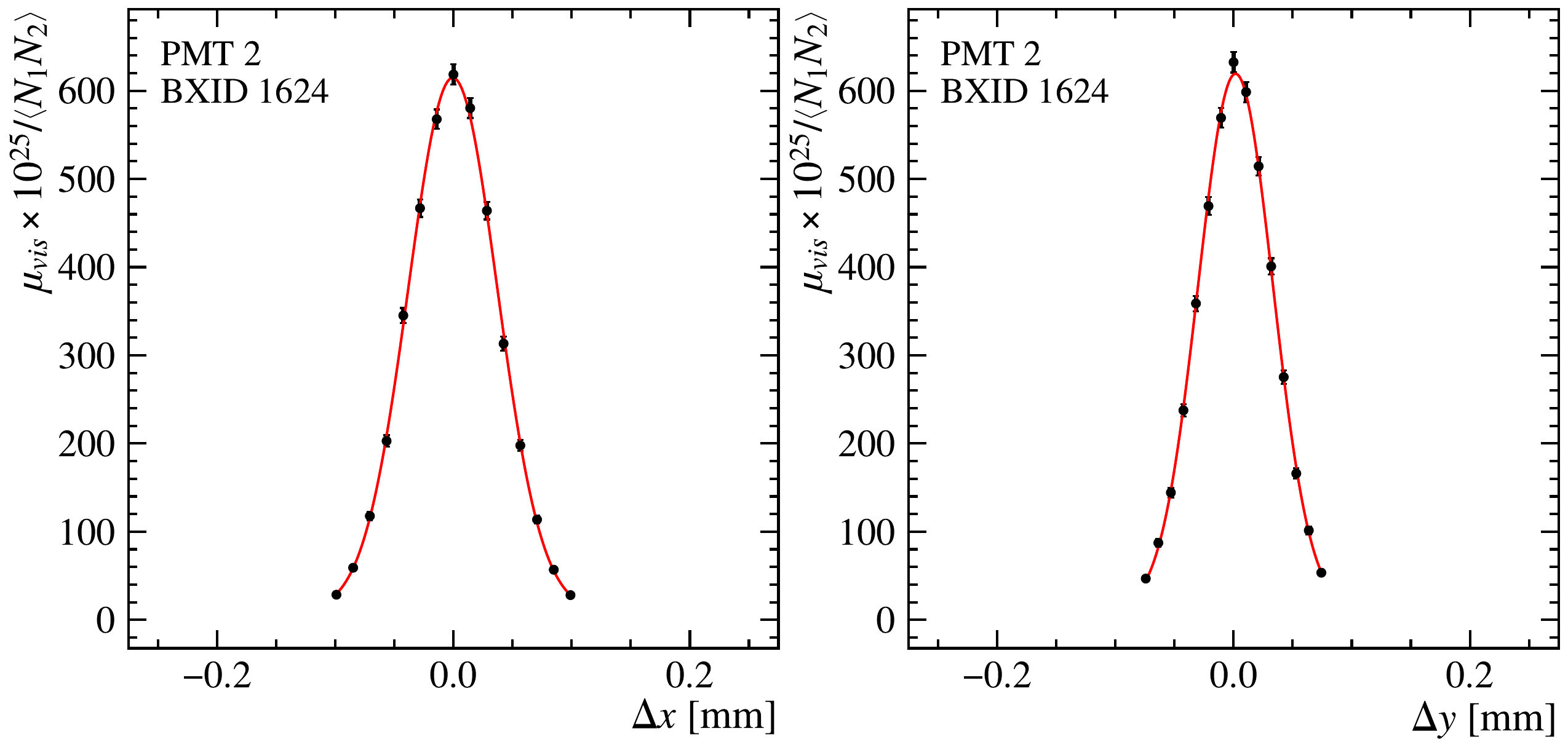}
    \caption{Fits to a 1-dimensional ${\rm OO}$ emittance scan, LHC fill 10802.}
    \label{fig:OO_scan_fits}
\end{figure}

\begin{figure}
    \centering
    \includegraphics[width=\linewidth]{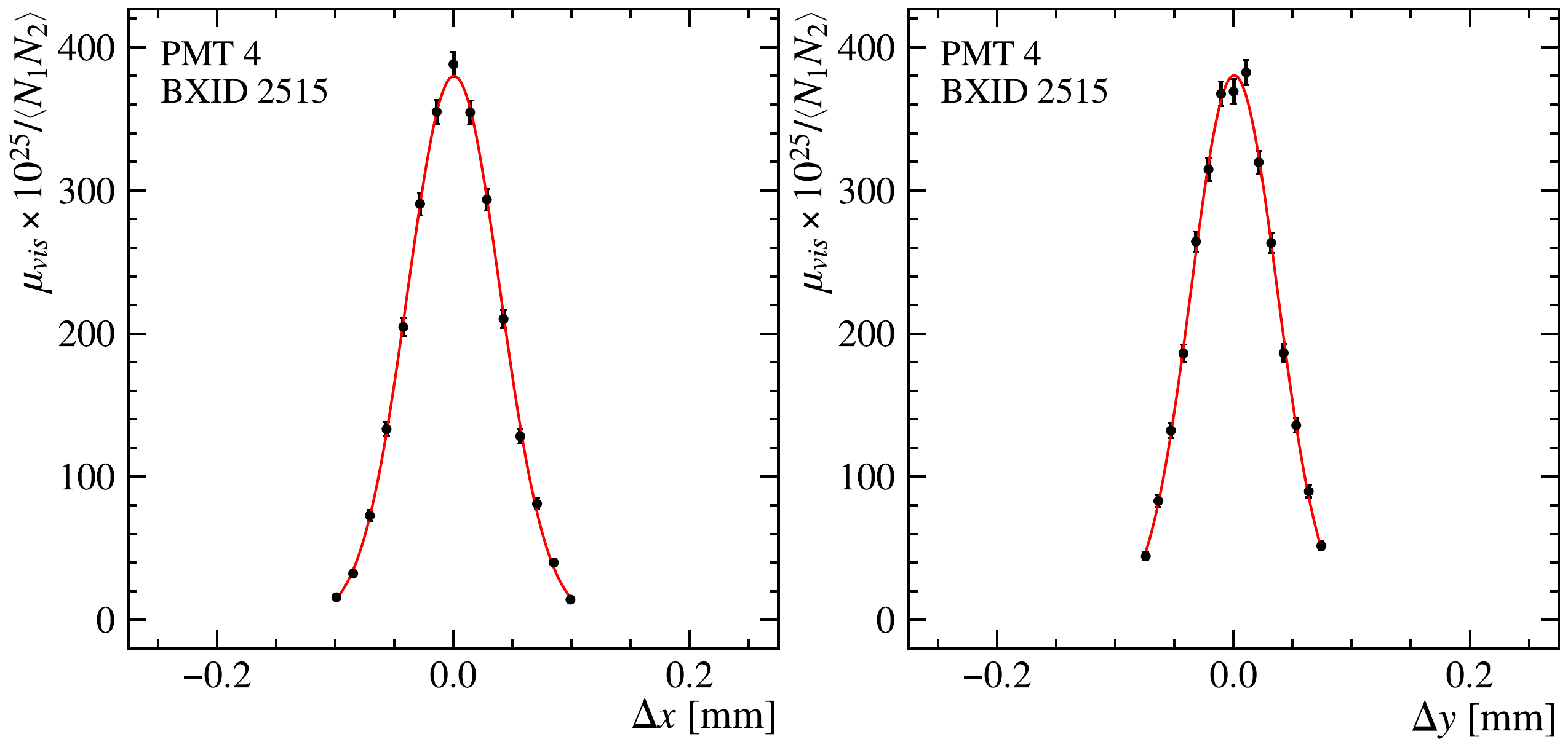}
    \caption{Fits to a 1-dimensional ${\rm NeNe}$ emittance scan, LHC fill 10813.}
    \label{fig:NeNe_scan_fits}
\end{figure}

\begin{figure}
    \centering
    \includegraphics[width=\linewidth]{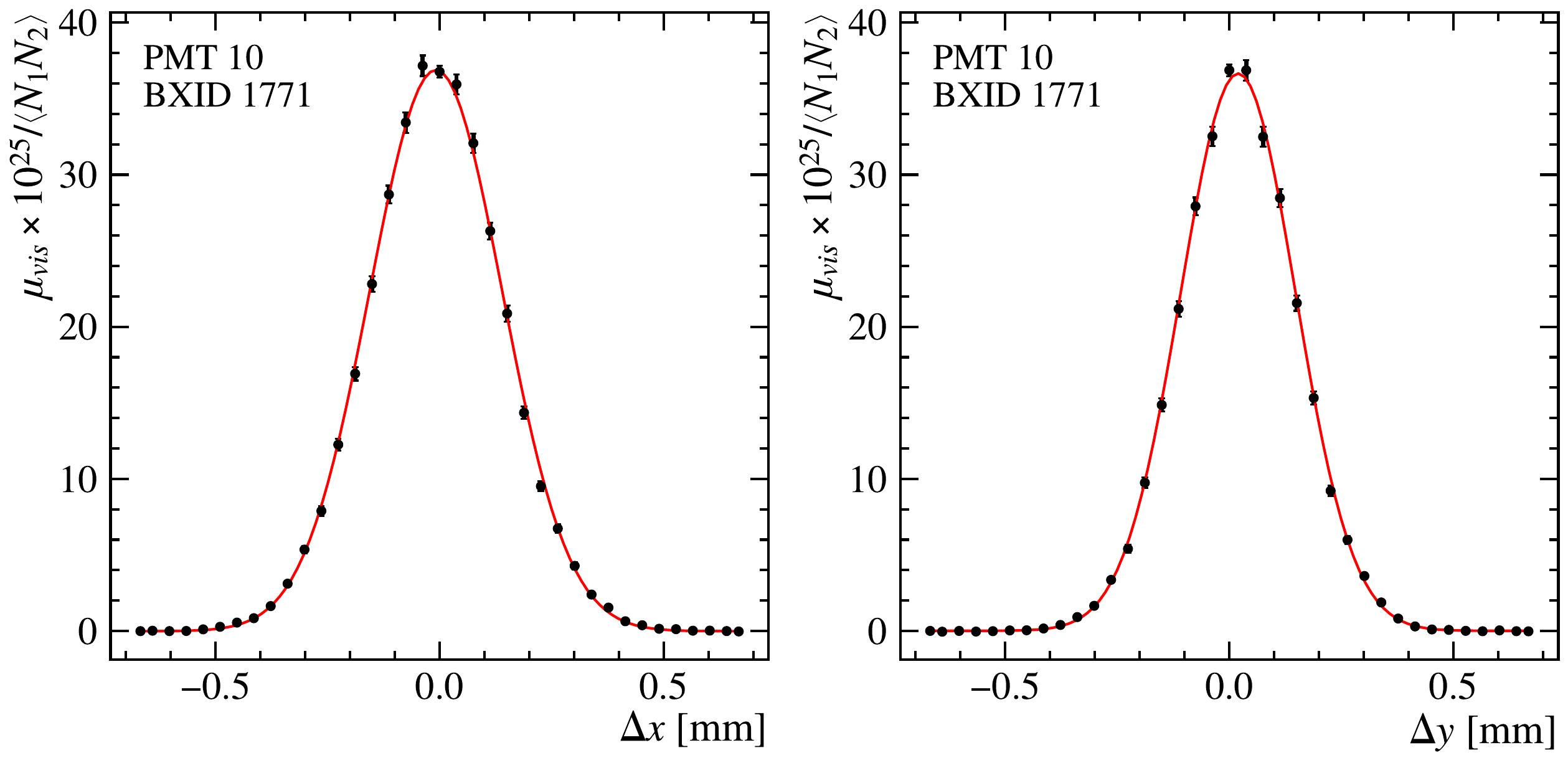}
    \caption{Fits to a 1-dimensional scan of a $\proton\proton$ vdM sequence, LHC fill 10821.}
    \label{fig:pp_scan_fits}
\end{figure}

\begin{figure}
    \centering
    \includegraphics[width=\linewidth]{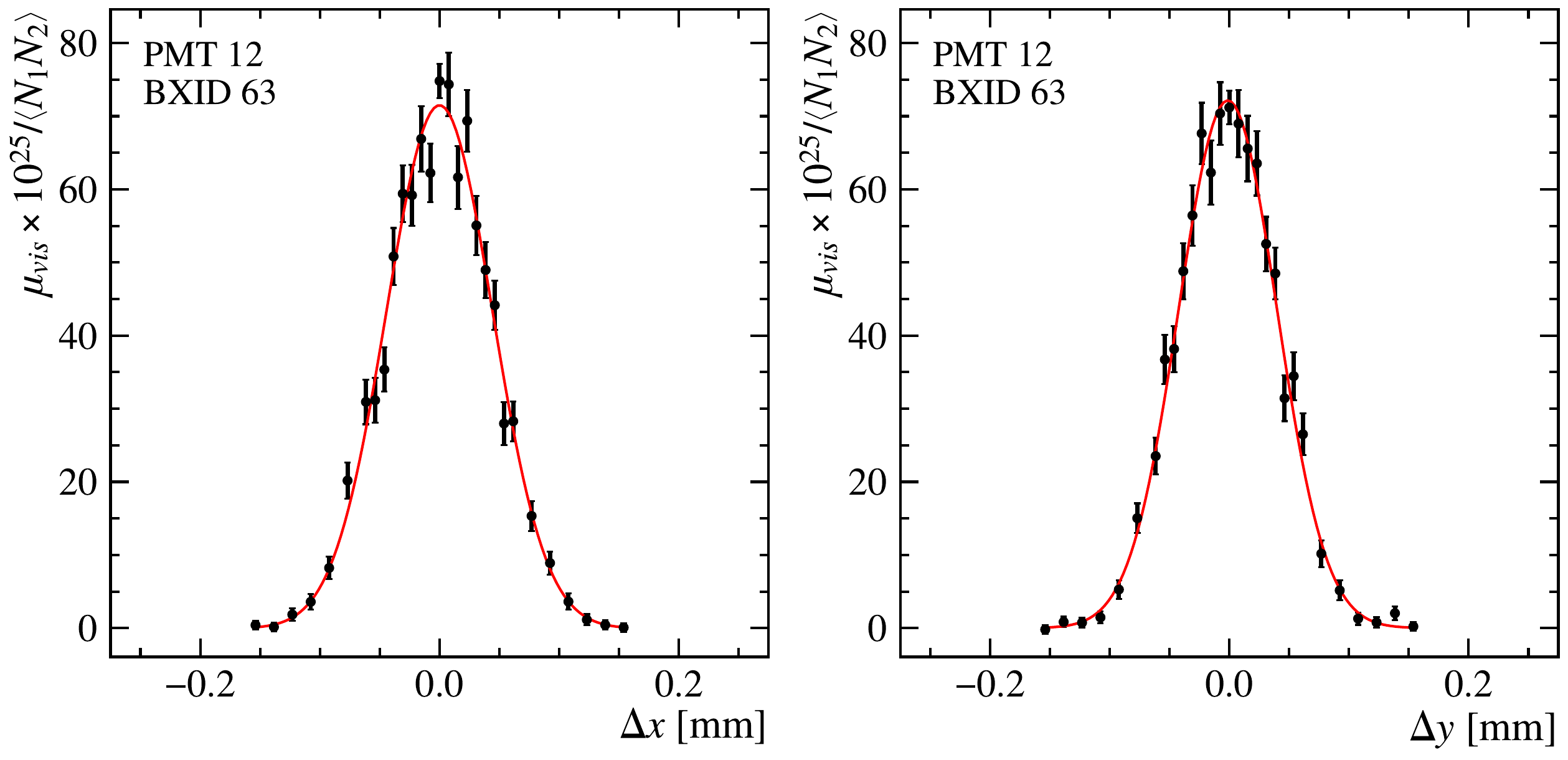}
    \caption{Fits to a 1-dimensional scan of a ${\rm PbPb}$ vdM sequence, LHC fill 11297.}
    \label{fig:PbPb_scan_fits}
\end{figure}

During vdM and emittance scans, all PLUME online values from the TELL40 readout are also saved in real time in ASCII format. 
This enables a rapid determination of the cross-sections, approximately within 30 minutes after the scans. 
This approach eliminates the need to wait for the second level of the High Level Trigger (\hlttwo) to process data files — a procedure that can take several days — and allows for a quick update of the reference cross-sections. 
This capability is particularly important during emittance scans, which are vdM scans performed over a reduced $x$ and $y$ range under nominal beam conditions. 
These scans are routinely performed to monitor the stability of the various detectors’ cross-sections over time. 
This feature was relevant during the light ion runs that were performed between June and July 2025 at the LHC, when the particle species and LHC running conditions frequently changed.
Indeed, during these runs, several emittance scans were performed prior to fills where stable beams were declared, to allow the LHC experiments to update their cross-sections for the $\proton\mathrm{O}$, $\mathrm{OO}$, and $\mathrm{NeNe}$ runs.
The ASCII files allowed for a fast determination of the visible cross-section, that were often employed just few hours after the emittance scans were performed. 
Some of the fits to data used to determine the visible cross-sections during 2025 are shown as an example in figures.~\labelcref{fig:pO_scan_fits,fig:OO_scan_fits,fig:NeNe_scan_fits,fig:pp_scan_fits,fig:PbPb_scan_fits} for $p\mathrm{O}$, $\mathrm{OO}$, $\mathrm{NeNe}$, $pp$, and $\mathrm{PbPb}$ collisions, respectively.

The data for each scan are fitted with a model consisting of a normal distribution and a constant offset accounting for residual uncorrected background. 
The probability density function for a scan in the crossing plane ($\Delta x$) can be written as \[ f(\Delta x;A_x,\Delta x_0,\sigma_x,C_x) = A_x\exp\left(-\frac{(\Delta x-\Delta x_0)^2}{2 \sigma_x^2}\right) + C_x, \] and similarly for the scan in the separation plane ($\Delta y$). 

The best-fit parameters $A_x$, $A_y$, $\sigma_x$  and $\sigma_y$ are combined together with the atomic numbers of the colliding beams, $Z_1Z_2$, to obtain the visible cross-section for each PMT, according to the formula \[ \sigmavis = 2\pi Z_1Z_2\sigma_x\sigma_y\frac{2A_xA_y}{A_x+A_y}, \] where the background contribution is neglected, as it is found to be compatible with zero within the uncertainties.
For illustration, the numbers involved in the computation of the visible cross-sections for the scans from figures~\labelcref{fig:pO_scan_fits,fig:OO_scan_fits,fig:NeNe_scan_fits,fig:pp_scan_fits,fig:PbPb_scan_fits} are reported in~\Cref{tab:scan_parameters}.
\begin{table}
    \caption{Results of the fits to the 1-dimensional scans shown in figures.~\labelcref{fig:pO_scan_fits,fig:OO_scan_fits,fig:NeNe_scan_fits,fig:pp_scan_fits,fig:PbPb_scan_fits}}
    \label{tab:scan_parameters}
    \centering
    \begin{tabular}{ccccccc}
    \toprule
    Scan type & $Z_1Z_2$ & $A_x$ $[10^{-25}]$ & $A_y$ $[10^{-25}]$ & $\sigma_x$ $[\si{\micro\m}]$  & $\sigma_y$ $[\si{\micro\m}]$  & $\sigmavis$ [mb]\\
    \midrule
    $p{\rm O}$ vdM & $8$ & $853 \pm 4$ & $853 \pm 5$ & $37.3 \pm 0.1$ & $31.1 \pm 0.1$ & $49.8 \pm 0.3$ \\
    ${\rm OO}$ emittance & $64$ & $605 \pm 6$ & $607 \pm 6$ & $37.6 \pm 0.4$ & $31.1 \pm 0.4$ & $285 \pm 5$ \\
    ${\rm NeNe}$ emittance & $100$ & $380 \pm 4$ & $379 \pm 6$ & $39.1 \pm 0.6$ & $36.4 \pm 0.9$ & $339 \pm 10$ \\
    $pp$ vdM & $1$ & $36.9 \pm 0.2$ & $36.6 \pm 0.2$ & $147.8 \pm 0.6$ & $128.2 \pm 0.6$ & $4.38 \pm 0.03$ \\
    ${\rm PbPb}$ vdM & $6724$ & $72 \pm 1$ & $72 \pm 1$ & $44.3 \pm 0.9$ & $41.0 \pm 0.8$ & $5509 \pm 168$ \\
    \bottomrule
    \end{tabular}
\end{table}

A vdM scan can be divided into separate sequences performed with different conditions, such as the extent of the displacement probed between the beams or the type of scan, \ie 1- or 2-dimensional. 
This implies that many equivalent sets of cross-sections can be extracted by analysing different sequences, each one possibly yielding a different value for the instantaneous luminosity; during data-taking, a random one has been chosen as the nominal set to be used for measuring the luminosity.

To assess a possible systematic uncertainty introduced by the arbitrary choice of the sequence to analyse, all possible sub-scans of the 2025 \proton\proton vdM scan have been processed to obtain the corresponding cross-sections, amounting to 18 different sets.
With these, the total integrated luminosity from a reference \proton\proton fill (namely 11221) has been recomputed to obtain 18 different values, which have been compared to the nominal one.
The largest relative deviation from the baseline integrated luminosity has been found to be 3.1\%.
This value is conservatively taken as the uncertainty associated with the choice of the reference cross-sections for the PLUME PMTs in the online luminosity measurement.

\subsubsection*{Luminosity computation}
\label{sec:lumi_determination}
From the background-subtracted interaction rates \mubkgsub and the corresponding
visible cross-sections, independent luminosity measurements are obtained for
each of the 44 PLUME PMTs.
These measurements are subsequently averaged to provide the final luminosity
value delivered to the LHC.
In addition, the \mubkgsub values are used to compute the luminosity for
individual bunch crossings, enabling bunch-by-bunch luminosity monitoring and
providing timely feedback to the LHC during operation.

The integrated luminosity recorded by the LHCb experiment in \(pp\) collisions at \(\sqrt{s}=13.6\)~\tev during the 2024--2026 data-taking period, as measured by PLUME, is found to be:
\begin{align*}
    \mathcal{L}_{2024} &= (\phantom{1}\LumiTwentyFour \pm  \LumiErrTwentyFour)\,\invfb, \nonumber \\
    \mathcal{L}_{2025} &= (\LumiTwentyFive \pm \LumiErrTwentyFive)\,\invfb, \nonumber \\
    \mathcal{L}_{2026} &= (\phantom{1}\LumiTwentySix \pm \LumiErrTwentySix)\,\invfb, \nonumber
\end{align*}
where the uncertainty is completely dominated by systematic effects, since the statistical uncertainty is negligible, of order $\mathcal{O}(10^{-4})$, given that a 2.4 s data-taking interval corresponds to a statistical uncertainty of 0.3\%.

Several sources of systematic uncertainty affect the online luminosity
determination with PLUME and are discussed below.
\begin{itemize}
    \item The first contribution originates from residual inaccuracies in the gain calibration procedure, which lead to deviations from the target PMT gain.
    Its impact on the instantaneous luminosity is evaluated using pseudo-experiments, in which the ADC distribution of each PMT is randomly shifted and the corresponding \mubkgsub value is recomputed.
    The RMS of the resulting luminosity distribution is taken as the systematic uncertainty associated with this effect, yielding a relative uncertainty of 0.4\% for all the data-taking periods.
    This corresponds to absolute uncertainties of 0.04, 0.05 and 0.02~\(\mathrm{fb}^{-1}\), respectively.
    \item A second contribution arises from differences among the sets of visible cross-sections measured with the PLUME online counters during the van der Meer scan campaigns.
    As discussed in Sec.~\ref{subsec:vdM}, this uncertainty is evaluated by comparing the luminosity obtained for a reference fill using different cross-section determinations.
    The largest relative uncertainty among the 2024, 2025, and 2026 data-taking periods, amounting to 3.1\%, is conservatively adopted for all three years. 
    This translates into absolute systematic uncertainties of 0.30, 0.37, and 0.17~\(\mathrm{fb}^{-1}\) for the 2024, 2025, and 2026 data-taking periods, respectively. 
    This contribution represents the dominant source of systematic uncertainty in the integrated luminosity determination.
    \item A systematic uncertainty associated with the use of the 44 independent measurements was evaluated by studying the distribution of the integrated-luminosity values obtained from the individual PMTs.
    A full \proton\proton data-taking year was used, and the systematic contribution was estimated from the RMS of the distribution after subtracting in quadrature the known contributions the gain-adjustment uncertainty (0.4\%) and the cross-section uncertainty (0.3\%).
    The resulting systematic uncertainty amounts to 1.7\%.
    \item Another source of systematic uncertainty is assessed through a comparison of the online luminosity measurements provided by the RICH1, RICH2, and VELO-Retina counters~\cite{LHCb:2025uci,LHCb-DP-2025-006} over the full dataset available for each data-taking period. 
    The largest relative spread observed among the integrated luminosities measured by the different counters across the 2024, 2025, and 2026 data-taking periods, amounting to 1.9\%, is conservatively adopted as the systematic uncertainty associated with the consistency of the online measurements. 
    This corresponds to absolute uncertainties of 0.18, 0.22, and 0.10~\invfb for the 2024, 2025, and 2026 data-taking periods, respectively.
\end{itemize}
A summary of the individual contributions for all data-taking periods is reported in Table~\ref{tab:plume_syst}. 
Since the same sources affect all years in the same way, the corresponding systematic uncertainties are taken to be identical and fully correlated across all data-taking periods.

\begin{table}[htbp]
    \centering
    \caption{Summary of the systematic uncertainties affecting the online luminosity determination with PLUME for the 2024--2026 data-taking periods. Absolute uncertainties for the individual years are given in \invfb, while the last column reports the corresponding relative uncertainties in percent.}
    \label{tab:plume_syst}
    \begin{tabular}{lcccc}
        \toprule
        Source of uncertainty &
        2024 &
        2025 &
        2026 &
        Relative \\
        & (\invfb) &
          (\invfb) &
          (\invfb) &
          (\%) \\
        \midrule
        Gain calibration
            & 0.04 & 0.05 & 0.02 & 0.4 \\
        Visible cross-section (vdM)
            & 0.30 & 0.37 & 0.17 & 3.1 \\
        Compatibility of measurements
            & 0.16 & 0.20 & 0.09 & 1.7 \\
        Online counters consistency
            & 0.18 & 0.22 & 0.10 & 1.9 \\
        \midrule
        Total (quadrature sum)
            & \LumiErrTwentyFour
            & \LumiErrTwentyFive
            & \LumiErrTwentySix
            & \LumiErrRelTwentyFour \\
        \bottomrule
    \end{tabular}
\end{table}

\subsection{Alternative online luminosity monitoring method}
\label{subsec:pmts_currents}
In addition to the primary luminosity determination based on interaction rates, PLUME implements an independent online luminosity monitoring approach intended to ensure operational continuity under non-nominal running conditions.

PLUME was able to provide online luminosity measurements for more than 99\% of the stable-beam time during the whole Run 3.
However, under specific conditions, such as temporary unavailability of the TFC signal or DAQ-related issues, an alternative system is required to replace the nominal PLUME luminosity determination.
This role is fulfilled by the Radiation Monitoring System (RMS)~\cite{Pugatch_2025}, which is based on robust and radiation-hard metal foil detector technology.
While highly reliable, the RMS has intrinsic limitations, most notably the inability to subtract the contribution from interactions occurring in the SMOG2 cell.

To complement the RMS and further enhance operational redundancy, an additional backup system based on the readout of PLUME PMT anode currents from the CAEN power supplies has been developed.
A dedicated ECS control script records the current drawn by each of the 48 PLUME PMTs at 3\,s intervals.
During the LHC RAMP and SQUEEZE phases~\cite{Evans_2008}, prior to the declaration of stable beams, the script accumulates these measurements and computes an average baseline current, which is subsequently subtracted from the total current.
Once stable beams are declared, 30 consecutive current measurements are used to derive a calibration factor for each PMT.
This calibration factor is determined once per fill and is kept constant for the entire duration of the fill, allowing the measured PMT currents to be converted into an online luminosity estimate.

\begin{figure}[!htbp]
\centering
\includegraphics[width=0.90\linewidth]{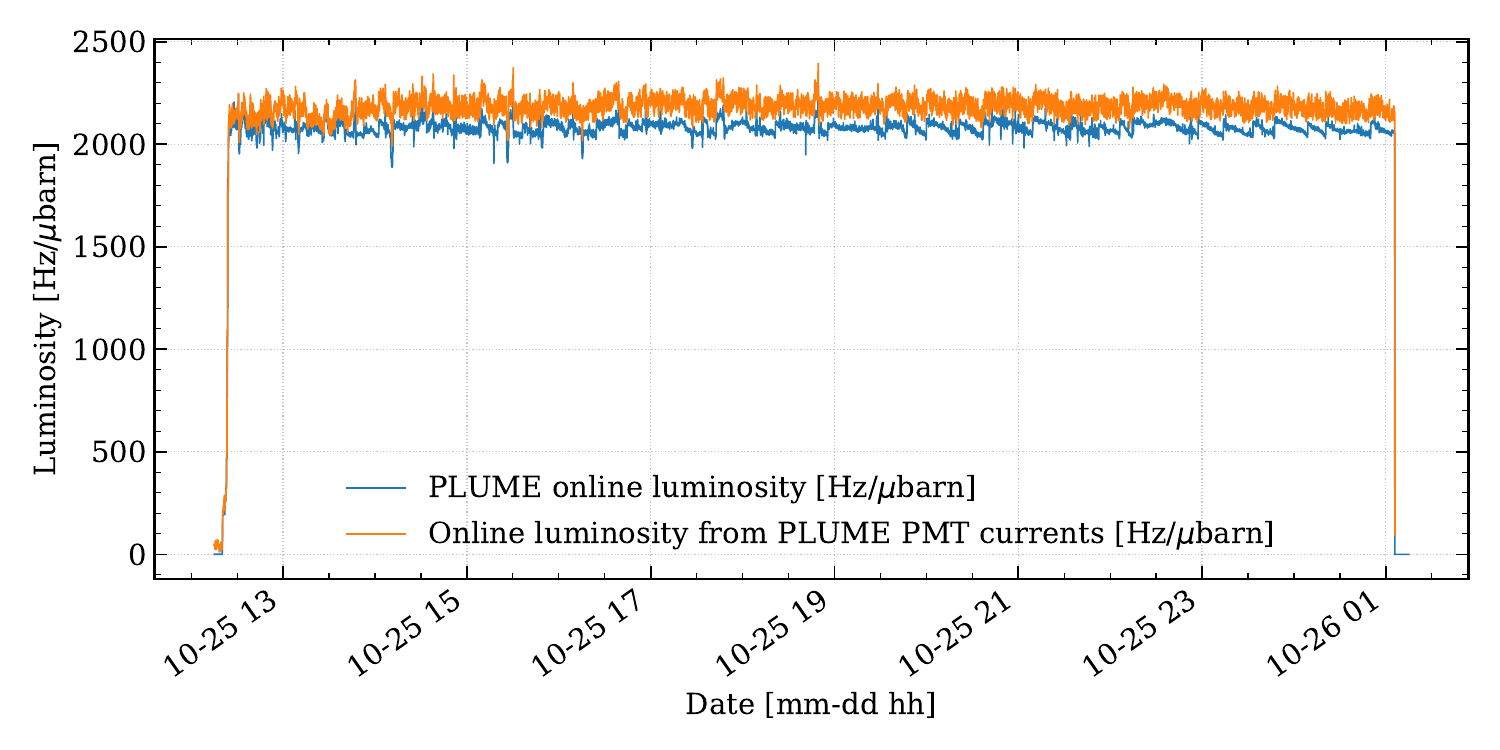} 
\caption{Comparison between (red) the nominal online luminosity and (light blue) backup luminosity from PMTs currents by PLUME as a function of time during a fill.} 
\label{fig:lumi_from_pmt_currents} 
\end{figure}

The system was deployed in mid-2025 and has been operating reliably since then.
Figure~\ref{fig:lumi_from_pmt_currents} shows a comparison between the nominal online luminosity determination and the backup measurement based on PMT currents.
A discrepancy at the level of 5.0\% is observed.
This difference is attributed to the calibration of the PMT-current-based system being performed in the absence of gas injection into the SMOG2 cell, whereas the nominal online luminosity measurement subtracts the contribution from SMOG2 interactions.

In principle, this contribution could be subtracted to improve the agreement with the nominal online measurement.
However, the determination of the SMOG2 contribution relies on the empty--beam rates measured by PLUME, which may themselves be affected under the same non-nominal operating conditions that motivate the use of the backup system.
For this reason, the SMOG2 correction is not applied.

Since the PMT-current-based luminosity measurement is used for less than 1\% of the total data-taking time, the impact of this residual disagreement on the integrated luminosity is negligible.
Considering the relative bias of 5\% during these periods, the resulting effect on the total integrated luminosity remains at the level of $5\times10^{-4}$.

One of the main limitations of the PMT-current-based method is the requirement of sufficient detector activity to produce measurable currents during operation.
As a consequence, this approach is applicable during nominal $pp$ running but not during PbPb or light-ion operation.
This limitation can be mitigated by dedicating one of the high-gain PMTs, originally used for the measurement of the collision phase relative to the LHC clock, to online luminosity monitoring.
By operating this PMT at a gain higher than the nominal value, the sensitivity to low instantaneous luminosities is significantly enhanced.
This strategy has been successfully tested during $pp$ operation and is foreseen for deployment during future PbPb data-taking periods.

\subsection{Measurement of the collision phase relative to the LHC clock}
\label{subsec:beam_phase_timing}
\begin{figure}
    \centering
    \includegraphics[width=0.7\linewidth]{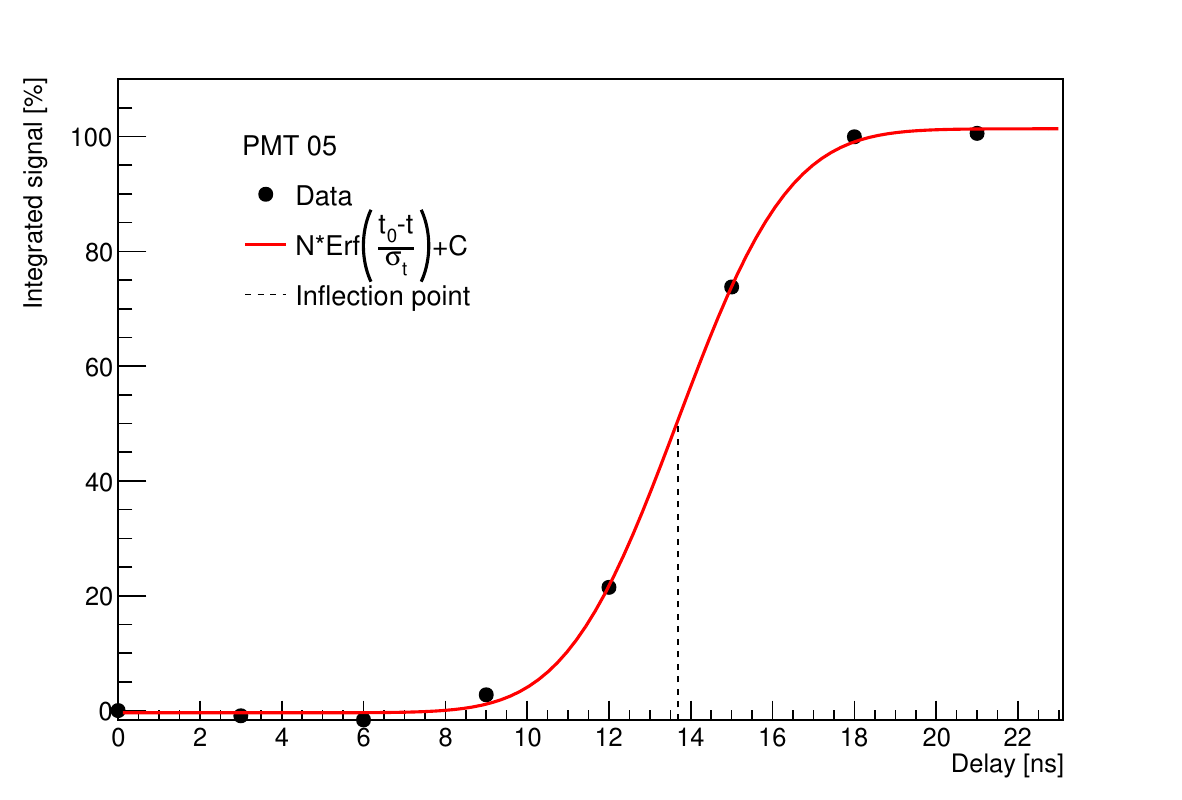}
    \caption{Example of a fit to the shape of the integral of the 8 copies of one timing PMT signal with increasing delays added. The data points are shown as black dots, the red curve is the best-fit function, and the dashed line corresponds to the inflection point.}
    \label{fig:s_shape_fit}
\end{figure}

The information from each PMT dedicated to the beam-phase timing is stored in 32-bit registers, which are read out using an ECS script.
These values are decoded and fitted with an error function to extract the inflection point, which is used as a timestamp.
One example of such fits is shown in figure~\ref{fig:s_shape_fit}.
The timestamps from the four PMTs are then averaged and used as a proxy for the collision phase relative to the LHC clock.
During the initial phase of operation, this measurement was performed at the HLT2 monitoring level and was therefore limited by the available statistics.
Following the implementation of the timing measurement directly in the front-end firmware, the available statistics increased significantly, leading to a more stable and precise determination of the collision--clock phase difference.

\begin{figure}[!htbp]
    \centering
    \includegraphics[width=0.9\linewidth]{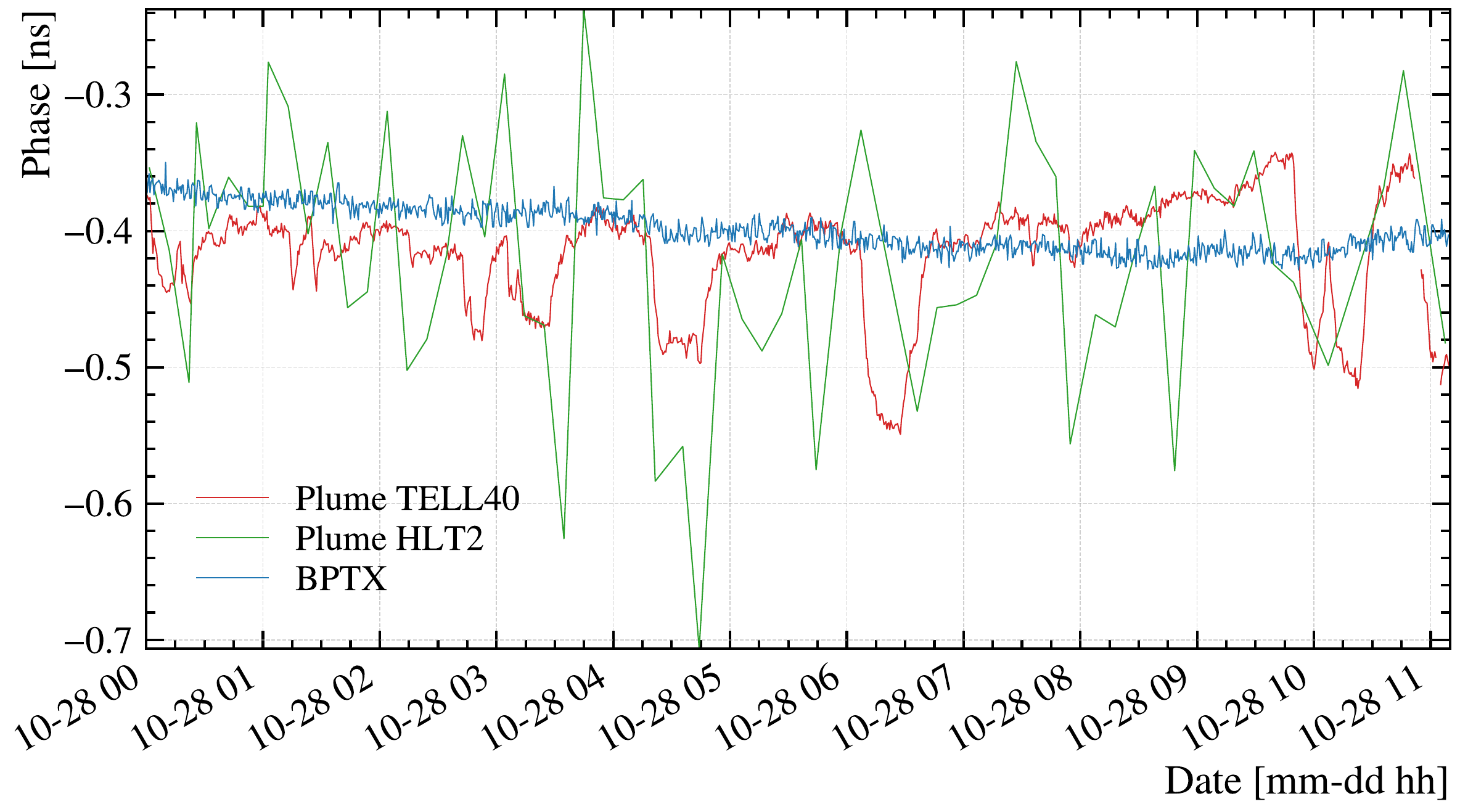} 
    \includegraphics[width=0.6\linewidth]{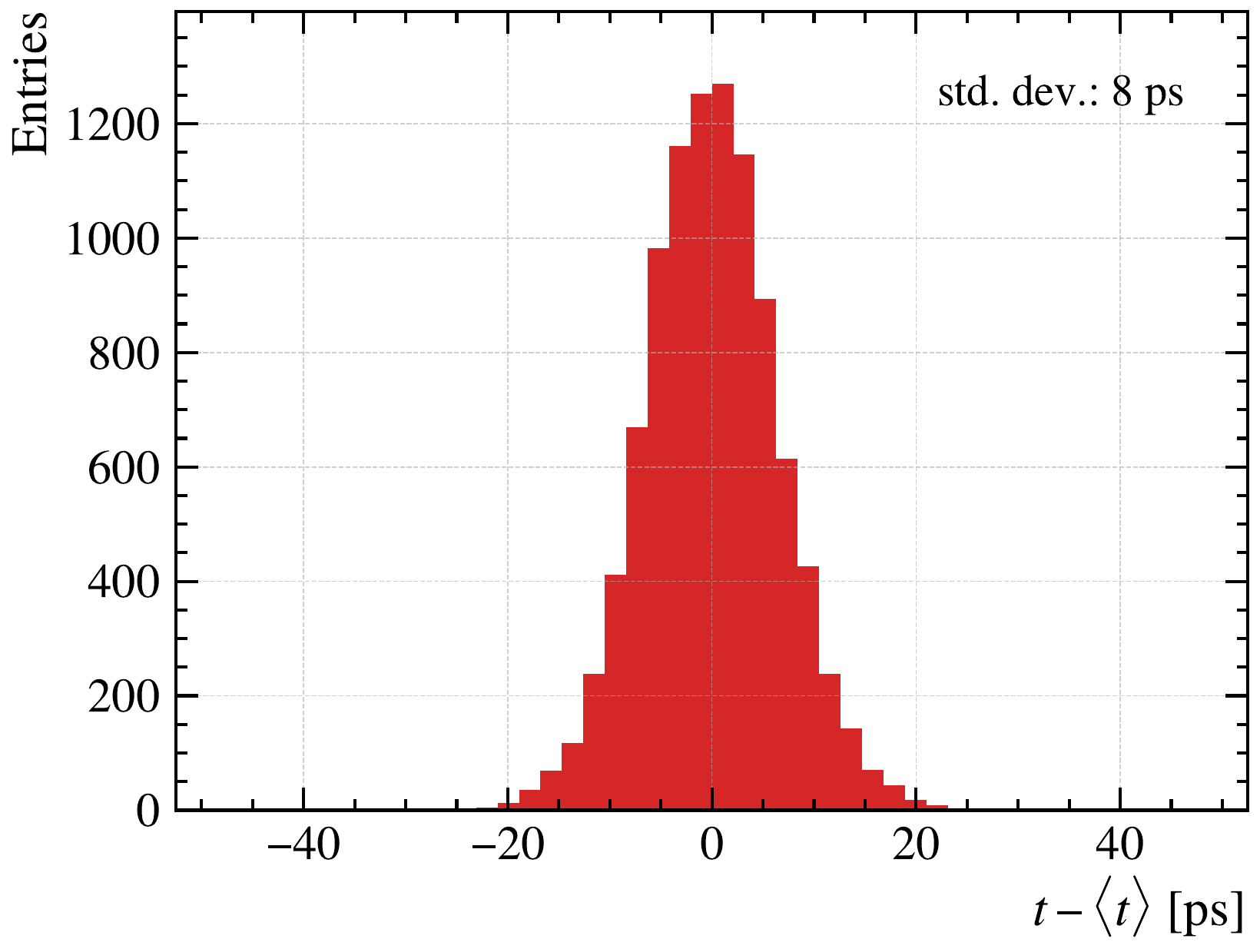} 
    \caption{(Top) Evolution of the LHC beam phase measurement performed by the BPTX (blue), by PLUME's HLT2 counters (green), and by PLUME's TELL40 data (red). The PLUME measurement are averaged in the same time-intervals as the BPTX measurements, to allow an easier comparison. (Bottom) Distribution of the PLUME's TELL40 measurement residuals for the fill represented in the top part of the figure, obtained from data sampled every 5 s after subtracting a rolling 60 s average to remove slow drifts. This procedure isolates the intrinsic resolution of the measurement by suppressing low-frequency variations that would otherwise broaden the distribution.}
    \label{fig:beam_phase_timing} 
\end{figure}

The top part of figure~\ref{fig:beam_phase_timing} shows the resulting measurements as a function of time, compared with the BPTX measurements.
The value measured at the HLT2 monitoring level is fully compatible with that obtained directly from the TELL40 firmware, although it exhibits larger statistical fluctuations due to the limited available statistics.

A good overall agreement is observed, although differences up to around 200\,ps are present in the time evolution of the measurements. 
These differences are attributed to the intrinsically different physical quantities measured by the two systems. 
The BPTX system measures the passage time of the individual LHC beams at a fixed longitudinal position, while PLUME measures the arrival time of particles produced in proton--proton collisions.
The origin of the dips and structures observed in the PLUME TELL40 measurements is not yet understood and is currently under investigation. 
Preliminary studies suggest that these features are not correlated with the position of the \proton\proton interaction vertices, the SMOG2 operating regime, or temperature variations in the LHCb experimental area. 
Further investigations are ongoing, with input from LHC accelerator experts, to better clarify their origin.

The bottom part of figure~\ref{fig:beam_phase_timing} shows the distribution of the PLUME’s TELL40 measurement residuals obtained from data sampled every 5 s, after subtracting a rolling 60 s average to remove slow drifts. The resolution achieved by the TELL40 measurement is about 8 ps.

In LHCb operation, the phase of the data-acquisition gate is adjusted whenever the measured collision--clock phase difference exceeds approximately 0.5\,ns.
This threshold is sufficient to ensure proper time alignment of the various detector subsystems.
Both the PLUME and BPTX measurements are capable of observing such phase shifts, thereby providing complementary and consistent inputs for the timing alignment of the experiment.

\section{Conclusions}
\label{sec:conclusions}
The PLUME detector has been successfully commissioned and operated during the LHC Run 3, providing online and offline luminosity measurements for the LHCb experiment.
The detector performance and operational stability have been demonstrated under routine running conditions, including periods of high luminosity and dedicated beam-based calibration campaigns.

A dedicated firmware implementation has been developed for the PLUME readout electronics to enable the online determination of interaction rates and the background-subtracted quantity \mubkgsub.
Following calibration with van der Meer scans, these measurements are converted into luminosity values on a per-bunch-crossing basis and integrated over time.
The online luminosity determination has been shown to be robust and reliable throughout data taking.

A detailed evaluation of the systematic uncertainties affecting the online luminosity measurement has been presented.
The dominant contribution arises from the determination of the visible cross-sections, while additional effects related to gain stability and the consistency of independent online luminosity measurements have been quantified.
By combining the individual contributions in quadrature, assuming them to be uncorrelated within a given data-taking period, and treating the corresponding relative uncertainties as fully correlated across the 2024, 2025, and 2026 data-taking periods, the total systematic uncertainty on the online luminosity determination is found to be \LumiErrTwentyFour~\invfb, \LumiErrTwentyFive~\invfb, and \LumiErrTwentySix~\invfb\ for 2024, 2025, and 2026, respectively. 
Under these assumptions, the total relative uncertainty is the same for all three years and amounts to \LumiErrRelTwentyFour\unskip\%.

Using this procedure, the integrated luminosity recorded by the LHCb experiment in \(pp\) collisions at \(\sqrt{s}=13.6\)~\tev and measured by the PLUME online counters for detector performance monitoring is determined to be
\(\mathcal{L}_{2024}= (\LumiTwentyFour \pm \LumiErrTwentyFour)\,\invfb\),
\(\mathcal{L}_{2025}= (\LumiTwentyFive \pm \LumiErrTwentyFive)\,\invfb\), and
\(\mathcal{L}_{2026}= (\LumiTwentySix \pm \LumiErrTwentySix)\,\invfb\)
for the 2024, 2025, and 2026 data-taking periods, respectively.
Considering the systematic uncertainties as fully correlated across the three years, the total integrated luminosity recorded by the LHCb experiment during the 2024--2026 data-taking period, as measured by the PLUME online counters, is determined to be
\(\mathcal{L} = (\LumiRunThree \pm \LumiErrRunThree)\,\invfb\).

The luminosity values used in physics analyses are determined separately through dedicated offline calibrations based on van der Meer scans and Beam Gas Imaging techniques, and will be reported in a dedicated publication.

In addition to luminosity measurements, PLUME has been able to provide a measurement of the collision phase with respect to the LHC clock.
The implementation of the timing measurement directly in firmware has enabled a high-statistics and stable determination of the collision--clock phase
difference.
The observed agreement with the BPTX measurements, together with the additional sensitivity of PLUME to longitudinal shifts of the interaction point, demonstrates the complementarity of the two approaches and the usefulness of PLUME for continuous timing alignment during LHCb operation.

Overall, the results presented in this paper establish PLUME as a reliable and versatile instrument for online luminosity monitoring and timing measurements
in LHCb during Run~3.